\documentclass[secnumarabic, nobibnotes, aps, prd]{revtex4-2}
\usepackage{graphicx}
\usepackage{amsmath}
\usepackage{subcaption}
\usepackage{multirow}
\usepackage{color}

\begin{document}

\title{Thermodynamic topology of topological black hole in $F(R)$-ModMax
gravity's rainbow}
\author{B. Eslam Panah$^{1,2,3}$\footnote{
email address: eslampanah@umz.ac.ir}, B. Hazarika$^{4}$\footnote{%
email address: bidyuthazarika1729@gmail.com}, and P. Phukon$^{4,5}$\footnote{%
email address: prabwal@dibru.ac.in}}
\affiliation{$^{1}$ Department of Theoretical Physics, Faculty of Basic Sciences, University of Mazandaran, P. O. Box 47416-95447, Babolsar, Iran\\
$^{2}$ ICRANet-Mazandaran, University of Mazandaran, P. O. Box 47416-95447, Babolsar, Iran\\
$^{3}$ ICRANet, Piazza della Repubblica 10, I-65122 Pescara, Italy\\
$^{4}$ Department of Physics, Dibrugarh University, 786004, Dibrugarh, Assam, India\\
$^{5}$ Theoretical Physics Division, Centre for Atmospheric Studies, Dibrugarh University, Dibrugarh, Assam, India}

\begin{abstract}
In order to include the effect of high energy and topological parameters on
black holes in $F(R)$ gravity, we consider two corrections to this gravity:
energy-dependent spacetime with different topological constants, and a
nonlinear electrodynamics field. In other words, we combine $F(R)$ gravity's
rainbow with ModMax nonlinear electrodynamics theory to see the effects of
high energy and topological parameters on the physics of black holes. For
this purpose, we first extract topological black hole solutions in $F(R)$%
-ModMax gravity's rainbow. Then, by considering black holes as thermodynamic
systems, we obtain thermodynamic quantities and check the first law of
thermodynamics. The effect of the topological parameter on the Hawking
temperature and the total mass of black holes is obvious. We also discuss
the thermodynamic topology of topological black holes in $F(R)$-ModMax
gravity's rainbow using the off-shell free energy method. In this formalism,
black holes are assumed to be equivalent to defects in their thermodynamic
spaces. For our analysis, we consider two different types of thermodynamic
ensembles. These are: fixed $q$ ensemble and fixed $\phi$ ensemble. We take
into account all the different types of curvature hypersurfaces that can be
constructed in these black holes. The local and global topology of these
black holes are studied by computing the topological charges at the defects
in their thermodynamic spaces. Finally, in accordance with their topological
charges, we classify the black holes into three topological classes with
total winding numbers corresponding to $-1, 0$, and $1$. We observe that the
topological classes of these black holes are dependent on the value of the
rainbow function, the sign of the scalar curvature, and the choice of
ensembles.
\end{abstract}

\maketitle

\section{\textbf{Introduction}}

From a cosmological perspective, the most critical issues in physics this
century are early-time inflation and late-time acceleration. To address
these problems, modified theories of Einstein's gravity as an alternative
approach have been introduced and studied in Refs. \cite%
{MOD1,MOD2,MOD3,MOD4,MOD5}. Among these modified theories of gravity, the $%
F(R)$ theory of gravity has gained considerable attention because it can
explain various phenomena. For example, the $F(R)$ theory of gravity can
describe the accelerating expansion of the universe today \cite%
{MOD1,MOD2,MOD3,MOD4,MOD5,Acc1,Acc2}, the inflation of the early universe 
\cite{Inflation1,Inflation2,Inflation3,Inflation4,Inflation5}, the existence
of dark matter \cite{DM1,DM2,DM3}, and massive compact objects \cite%
{MassC1,MassC2,MassC3,MassC4,MassC5,MassC6,MassC7,MassC8,MassC9}. Also, this
theory of gravity is in accordance with the predictions of the solar system 
\cite{Solar1,Solar2,Solar3, Solar4}. Furthermore, $F(R)$ gravity effectively
explains the full range of evolutionary epochs in the Universe and is
consistent with both the Newtonian and post-Newtonian approximations \cite%
{Newton1,Newton2}. It is notable that in the action of the $F(R)$ gravity
theory, $F(R)$ is an arbitrary function of the scalar curvature, $R$.

The evidence from observations predicts that black holes can absorb all
forms of matter, including charged matter, through gravitational collapse.
Therefore, it will be important to explore the existence of charged black
holes. According to the fact that the $F(R)$ theory of gravity was capable
of describing some phenomena in the context of cosmological and
astrophysical, it is crucial to examine black hole solutions in this theory
that are coupled with electromagnetic fields. Previous studies have already
investigated the solutions for charged black holes in $F(R)$ gravity, taking
into account both linear (Maxwell) and nonlinear electromagnetic fields \cite%
{BH1,BH2,BH3,BH4,BH5,BH6,BH7,BH8,BH9,BH10,BH11,BH12,BH13,BH14,BH15,BH16,BH17,BH18,BH19,BH20,BH21,BH22,BH23,BH24,BH25,BH26}%
. In this work, we are interested in to study topological charged black
holes in $F(R)$ gravity, which are coupled with a new model of nonlinear
electromagnetic field.

Maxwell's theory of electrodynamics is a remarkable theory that effectively
explains many phenomena in classical electrodynamics. However, when it comes
to high-energy physics, this theory faces challenges and fails to resolve
the singularity of the electric field at the origin of the electrical
charge. To address these issues, nonlinear electrodynamics (NED) was
introduced as an alternative. NED offers several advantages over Maxwell's
theory. It can account for the self-interaction of virtual electron-positron
pairs \cite{Pair1,Pair2,Pair3}, the elimination of singularities associated
with black holes \cite{Regular0,Regular1,Regular2,Regular3}, and the Big
Bang \cite{BBSin1,BBSin2,BBSin3}, the modification of gravitational redshift
around super-strong magnetized compact objects \cite{MagNS1,MagNS2}, the
explanation of radiation propagation within specific substances \cite%
{Radi1,Radi2,Radi3,Radi4}, as well as the effects of the NED field on
pulsars and highly magnetized neutron stars \cite{NSI,NSII}. Born and Infeld
proposed the initial concept of NED in 1934 \cite{BI}, and it successfully
addresses several issues encountered in Maxwell's theory, including the
removal of the electric field's singularity at the center of point
particles. Additionally, there are other NED theories such as Power-Maxwell
NED (in which the Lagrange function is an arbitrary power of Maxwell's
Lagrange function) \cite{PM1,PM2,PM3,PM4,PM5}, as well as Euler-Heisenberg 
\cite{Heisenberg}, logarithmic \cite{Log}, exponential \cite{Exp},
double-logarithmic \cite{dLog}, arcsin \cite{arcsin}, and other forms of NED
have been introduced in Refs. \cite{NED1,NED2,NED3,NED4,NED5}. On the other
hand, Maxwell electrodynamics is particularly noteworthy for its duality and
scale invariance. In this regard and considering the importance of the NED,
Bandos, Lechner, Sorokin, and Townsend proposed a new model of NED known as
the modified Maxwell (ModMax) theory of nonlinear duality-invariant
conformal electrodynamics. The ModMax model of NED exhibits both duality and
conformal symmetries, similar to Maxwell's theory \cite{ModMaxI,ModMaxII}.

One of the open problems in physics is how to unify quantum mechanics and
general relativity together. There have been several proposed approaches to
address this challenge, such as string theory \cite{String}, loop quantum
gravity \cite{loop}, and spacetime foam models \cite{Foam}. These models all
have in common the concept of a minimum observable length, called the Planck
length. In some of these models, Lorentz invariance is violated by a
modified dispersion relation \cite{Ali}. This modified dispersion relation
introduces the Planck energy as a second relativistic invariant, alongside
the speed of light ($c$). By considering this modification, the special
theory of relativity can be seen as the classical limit of a more general
framework known as doubly special relativity (DSR) \cite{DSR}. This approach
has been successful in explaining anomalies observed in $TeV$ photons \cite%
{TeV}, and ultra-high-energy cosmic rays \cite{HighRay}.

In DSR, the dispersion relation may be written as 
\begin{equation}
E^{2}f^{2}(\varepsilon )-p^{2}g^{2}(\varepsilon )=m^{2}
\end{equation}%
where $\varepsilon =\frac{E}{E_{P}}$. In addition, $f(\varepsilon )$ and $%
g(\varepsilon )$, are rainbow functions, and have phenomenological
motivations. Also, $E$ and $E_{P}$, are the energy of the particle used to
analyze the spacetime, and the Planck energy, respectively. It is notable
that, in the limit $\underset{\varepsilon \rightarrow 0}{\lim }%
f(\varepsilon)=1$, and $\underset{\varepsilon \rightarrow 0}{\lim }%
g(\varepsilon )=1$ (known as infrared limit), modified dispersion relation
reduces to the standard energy dispersion relations. The generalization of
DSR to include curvature in spacetime was proposed by Magueijo and Smolin 
\cite{RainbowG}, which is known as gravity's rainbow. In gravity's rainbow,
the spacetime is represented by a family of parameters in the metric that is
parameterized by $\varepsilon $, causing the spacetime geometry to depend on
the energy of the particle that is being used to test it, thus creating a
rainbow of metric. By considering gravity's rainbow, black hole solutions
and some of their physical properties have been studied by numerous
researchers \cite%
{RainBH1,RainBH2,RainBH3,RainBH4,RainBH5,RainBH6,RainBH7,RainBH8,RainBH9,RainBH10,RainBH11,RainBH12,RainBH13,RainBH14,RainBH15,RainBH16,RainBH17,RainBH18,RainBH19,RainBH20}%
. In this paper, we are interested in to study topological $F(R)$%
-ModMax-black holes in gravity's rainbow.

A new concept in black hole thermodynamics recently introduced is the notion
of thermodynamic topology. Since its conceptualization \cite{28,29}, it is
expanded into several notable works in Refs. \cite%
{30,31,32,33,34,35,36,37,38,39,40,41,42,43,44,45,46,47,48,50,51,52,53,54,55,56,57,58,59,60,61,62,63,64,65,66,67,68,69,70,71,72,73,74,75,76,77,78}%
. In this framework, black hole solutions are viewed as topological defects
within their thermodynamic space. Local and global topology can be analyzed
by calculating the winding numbers at these defects. Black holes are then
categorized based on their overall topological charge. Additionally, a black
hole's thermal stability is connected to the sign of its winding number. The
key aspect associated with thermodynamic topology is the concept of
topological defects and associated topological charges with that defect. In
Ref. \cite{29}, it was suggested that all physical black hole solutions can
be understood as the zero points of the tensor field $\gamma_{\mu \nu}$,
defined by the following equation 
\begin{equation}
\gamma_{\mu \nu}=G_{\mu \nu}-\frac{8 \pi G}{c^4} T_{\mu \nu}.
\end{equation}

In black hole thermodynamics, this concept is expanded by creating a vector
field derived from the generalized off-shell free energy. The expression for
the off-shell free energy of a black hole with arbitrary mass given by \cite%
{29} 
\begin{equation}
\mathcal{F}=E-\frac{S}{\tau },
\end{equation}%
in this equation, $E$ represents the energy, which is equivalent to the mass 
$M$, and $S$ represents the entropy of the black hole. The time scale
parameter $\tau$ is allowed to vary freely. To utilize this generalized free
energy, a vector field is defined as follows \cite{29} 
\begin{equation}
\phi=\left(\phi^r,\phi^\Theta \right)=\left(\frac{\partial\mathcal{F}}{%
\partial r_{+}},-\cot\Theta ~\csc\Theta \right),  \label{phifield}
\end{equation}
by taking the first component of the $\phi$ field 
\begin{equation}
\frac{\partial \mathcal{F}}{\partial r_+}=\frac{\frac{\partial M}{\partial S}%
}{\frac{\partial r_+}{\partial S}}-\frac{1}{\tau}\frac{\partial S}{\partial
r_+} \frac{\partial \mathcal{F}}{\partial r_+}=T \frac{\partial S}{\partial
r_+}-\frac{1}{\tau}\frac{\partial S}{\partial r_+},
\end{equation}
where $\frac{\partial M}{\partial S}=T$ is the equilibrium temperature. To
find the zero point of the vector field, we set $\frac{\partial \mathcal{F}}{%
\partial r_+}=0$, thus the zero point condition comes out to be 
\begin{equation}
\tau=1/T,
\end{equation}
so, $\tau$ can be understood as the inverse of the equilibrium temperature ($%
T$) of the cavity surrounding the black hole. Finally, considering both the $%
\phi^r$ and $\phi^\Theta$ components, we find that the zero point of $\phi$
occurs at $\theta=\pi/2$ and $\tau=1/T$. This confirms that a black hole
solution is a zero point or a defect of the vector $\phi$. Consequently,
each black hole solution can be assigned a topological charge. We determine
the associated topological charge using Duan's $\phi$ mapping technique.
Detailed methods are outlined in section \ref{tt}. In section \ref{tt}, we
explore the thermodynamic topology of topological black holes within the
framework of $F(R)$-ModMax gravity's rainbow, using the off-shell free
energy approach. The analysis considers two distinct thermodynamic
ensembles: the fixed $q$ ensemble and the fixed $\phi$ ensemble. We examine
all possible types of curvature hypersurfaces that can be constructed within
these black holes. By calculating the topological charges at the defects in
their thermodynamic spaces, we study both the local and global topology of
these black holes. Finally, we classify the black holes into various
topological classes based on their topological charges. We also observe how
the topological classes of these black holes are influenced by the value of
the thermodynamic parameter of a particular ensemble.

\section{\textbf{F(R)-ModMax gravity and black hole solutions}}

\label{sec1} 

The action of $F(R)$ gravity coupled with ModMax nonlinear electrodynamics
field is given by 
\begin{equation}
\mathcal{I}_{F(R)}=\frac{1}{16\pi }\int_{\partial \mathcal{M}}d^{4}x\sqrt{-g}%
\left[ F(R)-4\mathcal{L}\right] ,  \label{actionF(R)}
\end{equation}%
where $g=det(g_{\mu \nu })$ is the determinant of metric tensor $g_{\mu \nu }
$. In the above action, $F(R)=R+f\left( R\right) $. $R$ and $f\left(
R\right) $, respectively, devote to scalar curvature and a function of
scalar curvature. Here, we consider $G=c=1$, where $G$ is the Newtonian
gravitational constant and $c$ is the speed of light. $\mathcal{L}$ is
related to the ModMax Lagrangian. The ModMax Lagrangian is defined in the
following form \cite{ModMaxI,ModMaxII} 
\begin{equation}
\mathcal{L}=\mathcal{S}\cosh \gamma -\sqrt{\mathcal{S}^{2}+\mathcal{P}^{2}}%
\sinh \gamma ,  \label{ModMaxL}
\end{equation}%
where $\gamma $ is a dimensionless parameter. Here, we called $\gamma $ as
the parameter of ModMax theory. In the ModMax Lagrangian, $\mathcal{S}$ is a
true scalar, and $\mathcal{P}$ is a pseudoscalar. They are defined as 
\begin{equation}
\mathcal{S}=\frac{\mathcal{F}}{4},~~~\&~~~\mathcal{P}=\frac{\widetilde{%
\mathcal{F}}}{4},
\end{equation}%
where $\mathcal{F}=F_{\mu \nu }F^{\mu \nu }$ is the Maxwell invariant. Also, 
$F_{\mu \nu }$ is the electromagnetic tensor and defined as $F_{\mu \nu
}=\partial _{\mu }A_{\nu }-\partial _{\nu }A_{\mu }$ (where $A_{\mu }$ is
the gauge potential). Besides, $\widetilde{\mathcal{F}}$ is defined 
\begin{equation}
\widetilde{\mathcal{F}}=F_{\mu \nu }\widetilde{F}^{\mu \nu },
\end{equation}%
and $\widetilde{F}^{\mu \nu }=\frac{1}{2}\epsilon _{\mu \nu }^{~~~\rho
\lambda }F_{\rho \lambda }$. Notably, the ModMax theory reduces to Maxwell's
theory, when $\gamma =0$.

In this work, we want to obtain electrically charged case. Therefore, we
have to omit $\mathcal{P}$, i.e., we consider $\mathcal{P}=0$ in the above
equations. By considering $\mathcal{P}=0$, the equations of motion of $F(R)$
-ModMax gravity, turns to 
\begin{eqnarray}
R_{\mu \nu }\left( 1+f_{R}\right) -\frac{g_{\mu \nu }F(R)}{2}+\left( g_{\mu
\nu }\nabla ^{2}-\nabla _{\mu }\nabla _{\nu }\right) f_{R} &=&8\pi \mathrm{T}%
_{\mu \nu },  \label{EqF(R)1} \\
&&  \notag \\
\partial _{\mu }\left( \sqrt{-g}\widetilde{E}^{\mu \nu }\right) &=&0,
\label{EqF(R)2}
\end{eqnarray}%
where $f_{R}=\frac{df(R)}{dR}$. In addition, $\mathrm{T}_{\mu \nu }$ is the
energy-momentum tensor with the following form 
\begin{equation}
4\pi \mathrm{T}^{\mu \nu }=\left( F^{\mu \sigma }F_{~~\sigma }^{\nu
}e^{-\gamma }\right) -e^{-\gamma }\mathcal{S}g^{\mu \nu },  \label{eq3}
\end{equation}%
also, $\widetilde{E}_{\mu \nu }$ in Eq. (\ref{EqF(R)2}), is given by 
\begin{equation}
\widetilde{E}_{\mu \nu }=\frac{\partial \mathcal{L}}{\partial F^{\mu \nu }}%
=2\left( \mathcal{L}_{\mathcal{S}}F_{\mu \nu }\right) ,  \label{eq3b}
\end{equation}%
in which $\mathcal{L}_{\mathcal{S}}=\frac{\partial \mathcal{L}}{\partial 
\mathcal{S}}$. Therefore, the ModMax field equation (i.e., Eq. (\ref{EqF(R)2}
)) for the electrically charged case leads to 
\begin{equation}
\partial _{\mu }\left( \sqrt{-g}e^{-\gamma }F^{\mu \nu }\right) =0.
\label{Maxwell Equation}
\end{equation}

Now, we want to create a topological four-dimensional static
energy-dependent spacetime. For this purpose, we follow the mentioned method
in Refs. \cite{RainbowG,PengW}, which is 
\begin{equation}
h\left( \varepsilon \right) =\eta ^{\mu \nu }e_{\mu }\left( \varepsilon
\right) \otimes e_{\nu }\left( \varepsilon \right) .
\end{equation}%
and 
\begin{equation}
e_{0}\left( \varepsilon \right) =\frac{1}{f\left( \varepsilon \right) }%
\widetilde{e_{0}}~,~~~\&~~~e_{i}\left( \varepsilon \right) =\frac{1}{g\left(
\varepsilon \right) }\widetilde{e_{i}},
\end{equation}%
where the tilde quantities (i.e. $\widetilde{e_{0}}$ and $\widetilde{e_{i}}$
) refer to the energy-independent frame fields. Using the above conditions,
we are able to create a suitable four-dimensional static spherical symmetry
energy-dependent spacetime for extracting electrically charged black holes
in $F(R)$-ModMax gravity's rainbow, which leads to the following form 
\begin{equation}
ds^{2}=-\frac{\psi (r)}{f^{2}(\varepsilon )}dt^{2}+\frac{1}{%
g^{2}(\varepsilon )}\left( \frac{dr^{2}}{\psi (r)}+r^{2}d\Omega
_{k}^{2}\right) ,  \label{Metric}
\end{equation}%
in which $\psi (r)$, $f(\varepsilon )$, and $g(\varepsilon )$ are the metric
function, and rainbow functions, respectively. Also, $d\Omega _{k}^{2}$ is 
\begin{equation}
d\Omega _{k}^{2}=\left\{ 
\begin{array}{ccc}
d\theta ^{2}+\sin ^{2}\theta d\varphi ^{2} &  & k=1 \\ 
d\theta ^{2}+d\varphi ^{2} &  & k=0 \\ 
d\theta ^{2}+\sinh ^{2}\theta d\varphi ^{2} &  & k=-1%
\end{array}%
\right. ,
\end{equation}%
where the constant $k$ shows that the boundary of $t=$ constant and $r=$
constant can be elliptic ($k=1$), flat ($k=0$) or hyperbolic ($k=-1$)
curvature hypersurface, and is known as topological constant.

The equations governing $F(R)$ gravity with a nonlinear matter field (Eq. (%
\ref{EqF(R)1})) are generally complex, making it challenging to obtain a
precise analytical solution. One approach to overcome this difficulty is to
consider the traceless energy-momentum tensor for the nonlinear matter
field. By doing so, it becomes possible to derive an exact analytical
solution from $F(R)$ gravity coupled with a nonlinear matter field.
Therefore, to obtain the solution for a black hole with constant curvature
in $F(R)$ theory of gravity coupled with the ModMax field, it is necessary
for the trace of the stress-energy tensor $T_{\mu \nu }$ to be zero \cite%
{Rcont1,Rcont2}. Assuming a constant scalar curvature $R=R_{0}=$ constant \cite{Cognola2005},
the trace of equation (\ref{EqF(R)1}) becomes 
\begin{equation}
R_{0}\left( 1+f_{R_{0}}\right) -2\left( R_{0}+f(R_{0})\right) =0,
\label{R00}
\end{equation}%
in which $f_{R_{0}}=$ $f_{R_{\left\vert _{R=R_{0}}\right. }}$.

Here, we solve the equation (\ref{R00}) in order to find $R_{0}$, which
leads to 
\begin{equation}
R_{0}=\frac{2f(R_{0})}{f_{R_{0}}-1}.  \label{R0}
\end{equation}

We find the $F(R)$-ModMax gravity's equations of motion by replacing Eq. (%
\ref{R0}) within Eq. (\ref{EqF(R)1}), which is 
\begin{equation}
R_{\mu \nu }\left( 1+f_{R_{0}}\right) -\frac{g_{\mu \nu }}{4}R_{0}\left(
1+f_{R_{0}}\right) =8\pi \mathrm{T}_{\mu \nu }.  \label{F(R)Trace}
\end{equation}

On the other hand, to obtain a radial electric field, we consider the
following gauge potential 
\begin{equation}
A_{\mu }=h\left( r\right) \delta _{\mu }^{t},  \label{A0}
\end{equation}%
by considering the mentioned gauge potential (\ref{A0}), the ModMax field
equation (\ref{Maxwell Equation}), and the mentioned energy-dependent
spacetime (\ref{Metric}), we obtain the following differential equation 
\begin{equation}
rh^{\prime \prime }(r)+2h^{\prime }(r)=0,  \label{hh}
\end{equation}%
in which the prime and double prime devotes to the first and second
derivatives of $r$, respectively. We can find a solution of the equation ( %
\ref{hh}), which is 
\begin{equation}
h(r)=-\frac{q}{r},  \label{hh1}
\end{equation}%
in the above equation, $q$ is an integration constant. This integration
constant is associated with the electric charge.

By taking into account the metric (\ref{Metric}), the derived $h(r)$ (\ref%
{Metric}), and the field equations (\ref{F(R)Trace}), we can derive the
subsequent set of differential equations 
\begin{eqnarray}
eq_{tt} &=&eq_{rr}=r\psi ^{\prime \prime }(r)+2\psi ^{\prime }(r)+\frac{%
rR_{0}}{2g^{2}\left( \varepsilon \right) }-\frac{2q^{2}e^{-\gamma
}f^{2}\left( \varepsilon \right) }{r^{3}\left( 1+f_{R_{0}}\right) },
\label{eq1} \\
&&  \notag \\
eq_{\theta \theta } &=&eq_{\varphi \varphi }=r\psi ^{\prime }(r)+\psi \left(
r\right) -k+\frac{r^{2}R_{0}}{4g^{2}\left( \varepsilon \right) }+\frac{%
q^{2}e^{-\gamma }f^{2}\left( \varepsilon \right) }{r^{2}\left(
1+f_{R_{0}}\right) },  \label{eq2}
\end{eqnarray}%
where the components of $tt$, $rr$, $\theta \theta $ and $\varphi \varphi $
of field equations (\ref{F(R)Trace}) are indicated by $eq_{tt}$, $eq_{rr}$, $%
eq_{\theta \theta }$ and $eq_{\varphi \varphi }$, respectively.

By using the differential equations (Eqs. (\ref{eq1}), and (\ref{eq1})), we
extract a precise solution for the constant scalar curvature (i.e. $R=R_{0}$
= constant). After performing several calculations, we can find the metric
function in the following form 
\begin{equation}
\psi (r)=k-\frac{m_{0}}{r}-\frac{R_{0}r^{2}}{12g^{2}\left( \varepsilon
\right) }+\frac{q^{2}f^{2}\left( \varepsilon \right) e^{-\gamma }}{\left(
1+f_{R_{0}}\right) r^{2}},  \label{g(r)F(R)}
\end{equation}%
where $m_{0}$ in the above equation is an integration constant, and this
constant of integration is connected to the black hole's geometric mass.
Besides, all of the field equations (\ref{F(R)Trace}) is satisfied by the
obtained solution (\ref{g(r)F(R)}). The obtained solution limit ourselves to 
$f_{R_{0}}\neq -1$ to have physical solution. In this solution ( \ref%
{g(r)F(R)}), the effects of ModMax's parameter, $F(R)$ gravity, and rainbow
function are clear. This solution can cover Reissner-Nordstr\"{o}m-(A)dS
black hole by considering suitable parameter. In other words, the solution (%
\ref{g(r)F(R)}) reduces to the following form 
\begin{equation}
\psi (r)=1-\frac{m_{0}}{r}-\frac{\Lambda r^{2}}{3}+\frac{q^{2}}{r^{2}},
\end{equation}%
when $f_{R_{0}}=0$, $R_{0}=4\Lambda $, $f^{2}\left( \varepsilon \right)
=g^{2}\left( \varepsilon \right) =1$ and $\gamma =0$.

To find the singularity of spacetime, we can study the Kretschmann scalar ($%
R_{\alpha \beta \gamma \delta }R^{\alpha \beta \gamma \delta }$). Indeed,
the Kretschmann scalar gives us information about the existence of the
singularity in spacetime. So, we calculate the Kretschmann scalar of the
energy-dependent spacetime (\ref{Metric}) 
\begin{equation}
R_{\alpha \beta \gamma \delta }R^{\alpha \beta \gamma \delta }=\frac{%
R_{0}^{2}}{6}+\frac{12m_{0}^{2}g^{4}\left( \varepsilon \right) }{r^{6}}-%
\frac{48m_{0}q^{2}e^{-\gamma }g^{4}\left( \varepsilon \right) f^{2}\left(
\varepsilon \right) }{\left( 1+f_{R_{0}}\right) r^{7}}+\frac{%
56q^{4}e^{-2\gamma }g^{4}\left( \varepsilon \right) f^{4}\left( \varepsilon
\right) }{\left( 1+f_{R_{0}}\right) ^{2}r^{8}},
\end{equation}%
the obtained Kretschmann scalar indicates that there is a curvature
singularity situated at the coordinate $r=0$, because $\lim_{r%
\longrightarrow 0}R_{\alpha \beta \gamma \delta }R^{\alpha \beta \gamma
\delta }\longrightarrow \infty $. Besides, we can see the effects of
ModMax's parameter, $F(R)$ gravity, and rainbow function in the Kretschmann
scalar. It is notable that by considering $\gamma \longrightarrow \infty $,
we can only remove the divergence of the electrical field. However, there is
a curvature singularity at $r=0$, because $\lim_{r\rightarrow 0}\left( \frac{%
12m_{0}^{2}g^{4}\left( \varepsilon \right) }{r^{6}}\right) \longrightarrow
\infty $. As a result, in the limit $\gamma \longrightarrow \infty $, the
energy-dependent spacetime includes a divergency at $r=0$, i.e, 
\begin{equation}
\lim_{r\rightarrow 0}R_{\alpha \beta \gamma \delta }R^{\alpha \beta \gamma
\delta }\longrightarrow \infty
\end{equation}%
when $\gamma \longrightarrow \infty $.

On the other hand, the asymptotical behavior of the Kretschmann scalar and
the metric function are given by 
\begin{eqnarray}
\lim_{r\longrightarrow \infty }R_{\alpha \beta \gamma \delta }R^{\alpha
\beta \gamma \delta } &\longrightarrow &\frac{R_{0}^{2}}{6},  \notag \\
&& \\
\lim_{r\longrightarrow \infty }\psi \left( r\right) &\longrightarrow &-\frac{%
R_{0}r^{2}}{12g^{2}\left( \varepsilon \right) },  \notag
\end{eqnarray}%
where reveal that the spacetime will be asymptotically (A)dS, when we
consider $R_{0}=4\Lambda $, and $\Lambda >0$ ($\Lambda <0$). Our finding
indicate that, the asymptotical behavior is independent of $\gamma $ and $k$%
. In other words, the parameter of ModMax and topological constant do not
affect the asymptotical behavior of the spacetime, but it depends on the
rainbow function $g\left( \varepsilon \right)$.

Now, we aim to determine the real roots of the metric function (\ref%
{g(r)F(R)}) to gather insights about the solution's inner and outer
horizons. Black holes, characterized by a curvature singularity at $r=0$,
generally possess at least one event horizon that conceals this singularity.
Nevertheless, it is worth noting that black holes, known as naked
singularities, lack an event horizon.

\begin{figure}[tbh]
\centering
\includegraphics[width=0.3\textwidth]{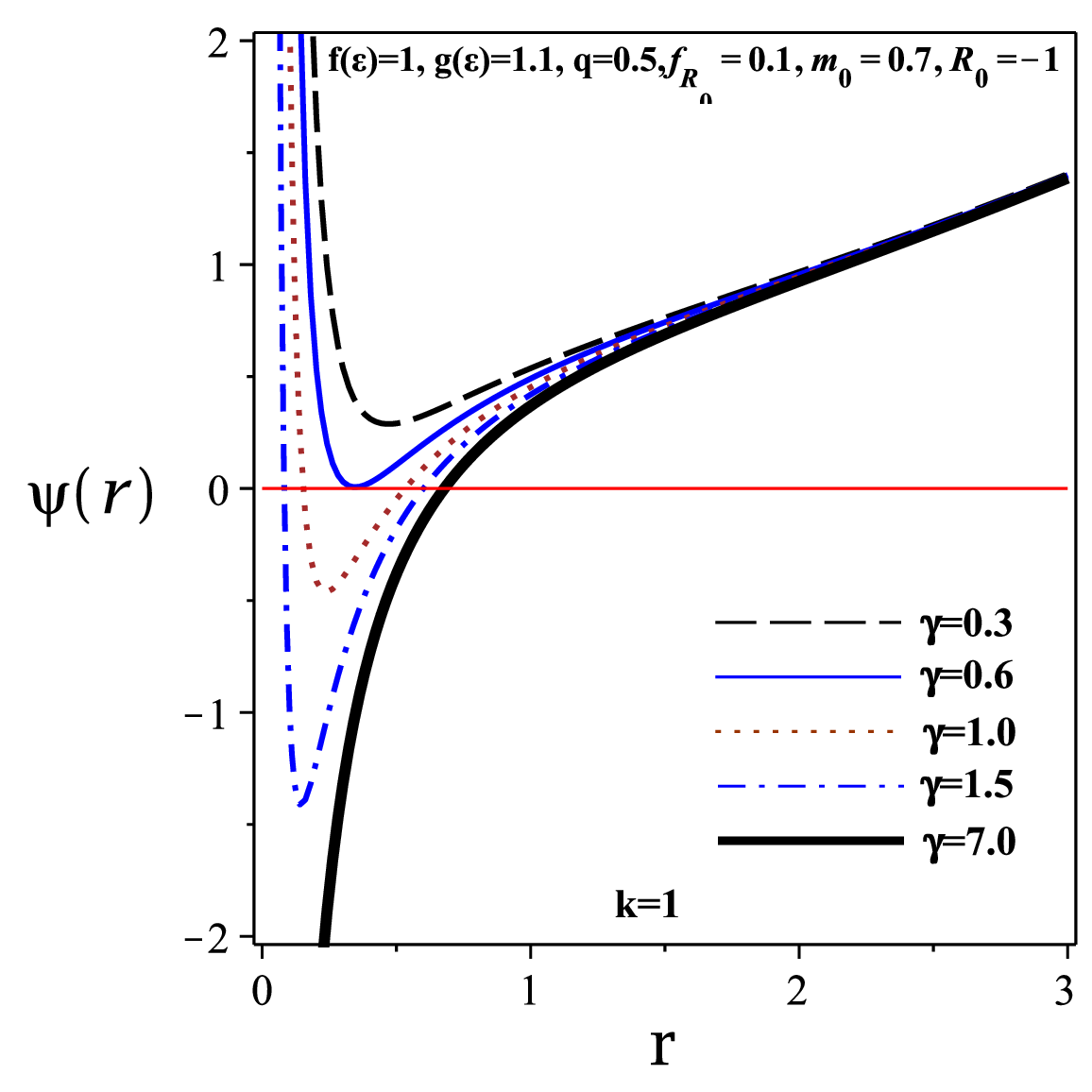} \includegraphics[width=0.3%
\textwidth]{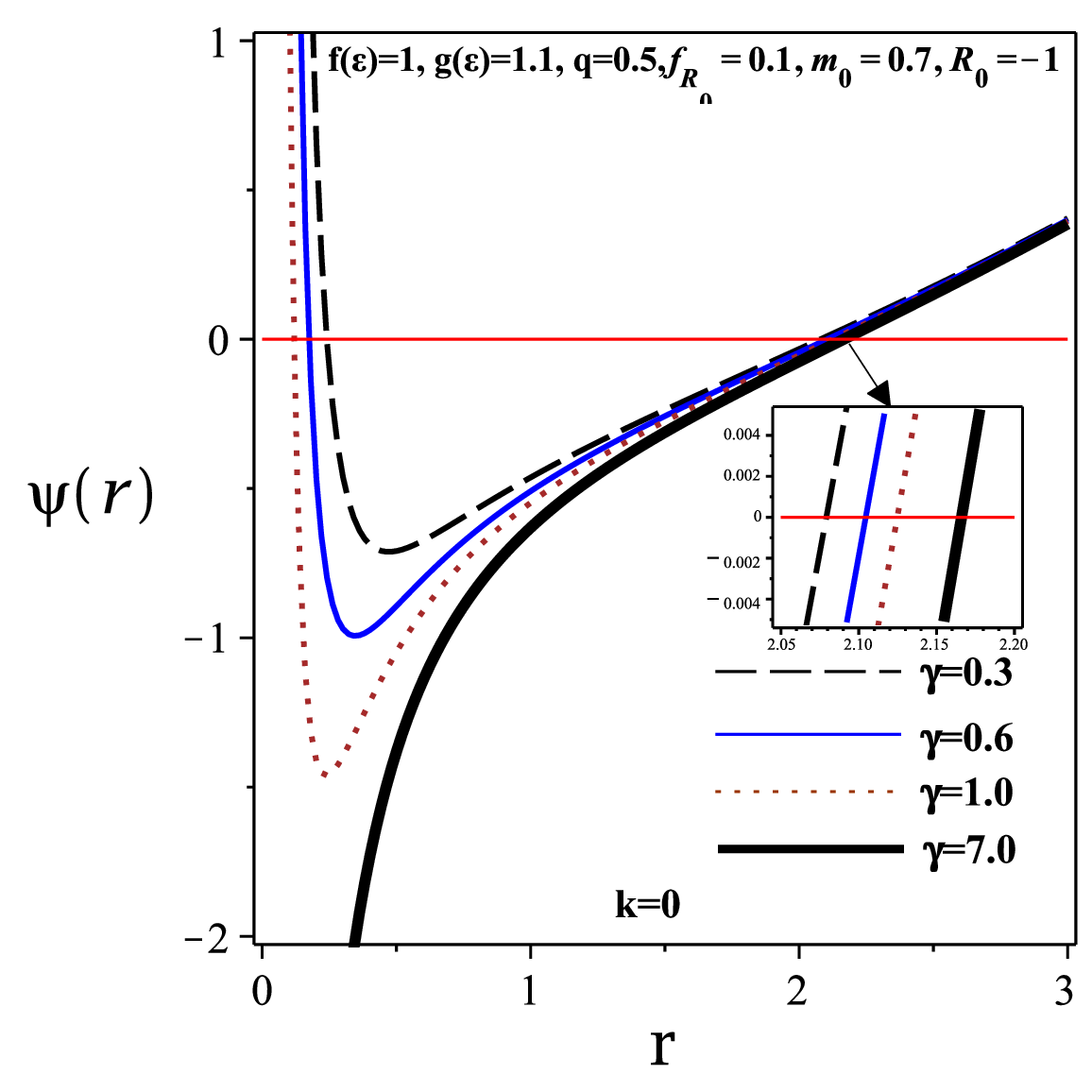} \includegraphics[width=0.3\textwidth]{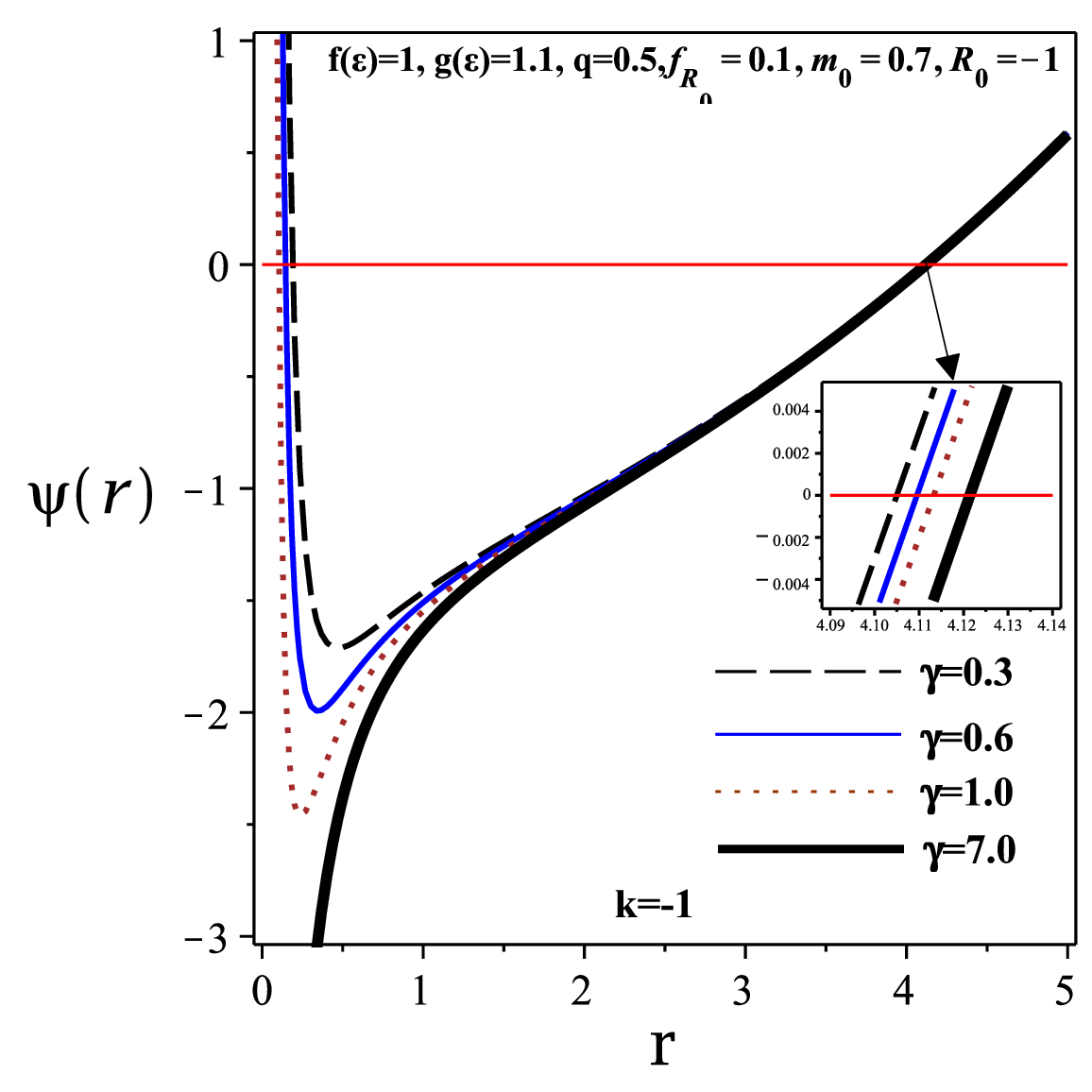} \newline
\caption{$\protect\psi(r)$ versus $r$ for different values of $k$, and $%
\protect\gamma$.}
\label{Fig1}
\end{figure}

We investigate the effects of the topological constant ($k$) and ModMax's
parameter ($\gamma $) on the roots of the metric function in Fig. \ref{Fig1}.

\textbf{For $k=1$:} By considering constant values for the parameters of the
studied system, there are two critical values for $\gamma$. Indeed, for $%
\gamma<\gamma_{critical_{1}}$, we find that the metric function has no
roots. This implies that we have what is called a naked black hole. For $%
\gamma=\gamma_{critical _{1}}$, we observe the appearance of one root (in
the extreme case). Considering $\gamma_{critical_{1}}<\gamma<%
\gamma_{critical_{2}}$, there are two roots for the metric function (as
shown in the left panel of Fig. \ref{Fig1}). It is important to note that
the outer root corresponds to the event horizon, and its size increases with
the ModMax parameter. In other words, as $\gamma$ increases, the black hole
becomes larger. Furthermore, the metric function has only one root when $%
\gamma > \gamma_{critical_{2}}$ (as shown in the left panel of Fig. \ref%
{Fig1}). It is worth mentioning that this root belongs to the event horizon.
However, the black hole continues to grow in size as $\gamma$ increases.

\textbf{For $k=0$:} Considering fixed values for the parameters of the
studied system (as before), there exists a $\gamma_{critical}$ such that the
black hole has two horizons when $\gamma<\gamma_{critical}$. When $\gamma$
is greater than $\gamma_{critical}$, the black hole has only one event
horizon. It is worth noting that as $\gamma$ increases, the event horizon
also increases, resulting in the presence of larger black holes (see the
middle panel in Fig. \ref{Fig1}).

\textbf{For $k=-1$:} The behavior for $k=-1$ is the same as for $k=0$.
Specifically, when $k=-1$, there is a critical value for $\gamma$ that
determines whether black holes have one or two roots. If $%
\gamma<\gamma_{critical}$, there are two roots, but if $\gamma>%
\gamma_{critical}$, there is only one root. Additionally, as the value of $%
\gamma$ increases, the event horizon expands, resulting in larger black
holes (see the right panel in Fig. \ref{Fig1}).

Our analysis of Figure. \ref{Fig1}, indicates that large black holes belong
to $k=-1$. Indeed, by considering the same values of parameters, and for
different values of $k$, the arrangement of the size of the radius of the
black holes follows as $r_{+}{}_{(k=-1)}>r_{+}{}_{(k=0)}>r_{+}{}_{(k=+1)}$,
where $r_{+}$ is the radius of the event horizon.

\begin{figure}[tbh]
\centering
\includegraphics[width=0.3\textwidth]{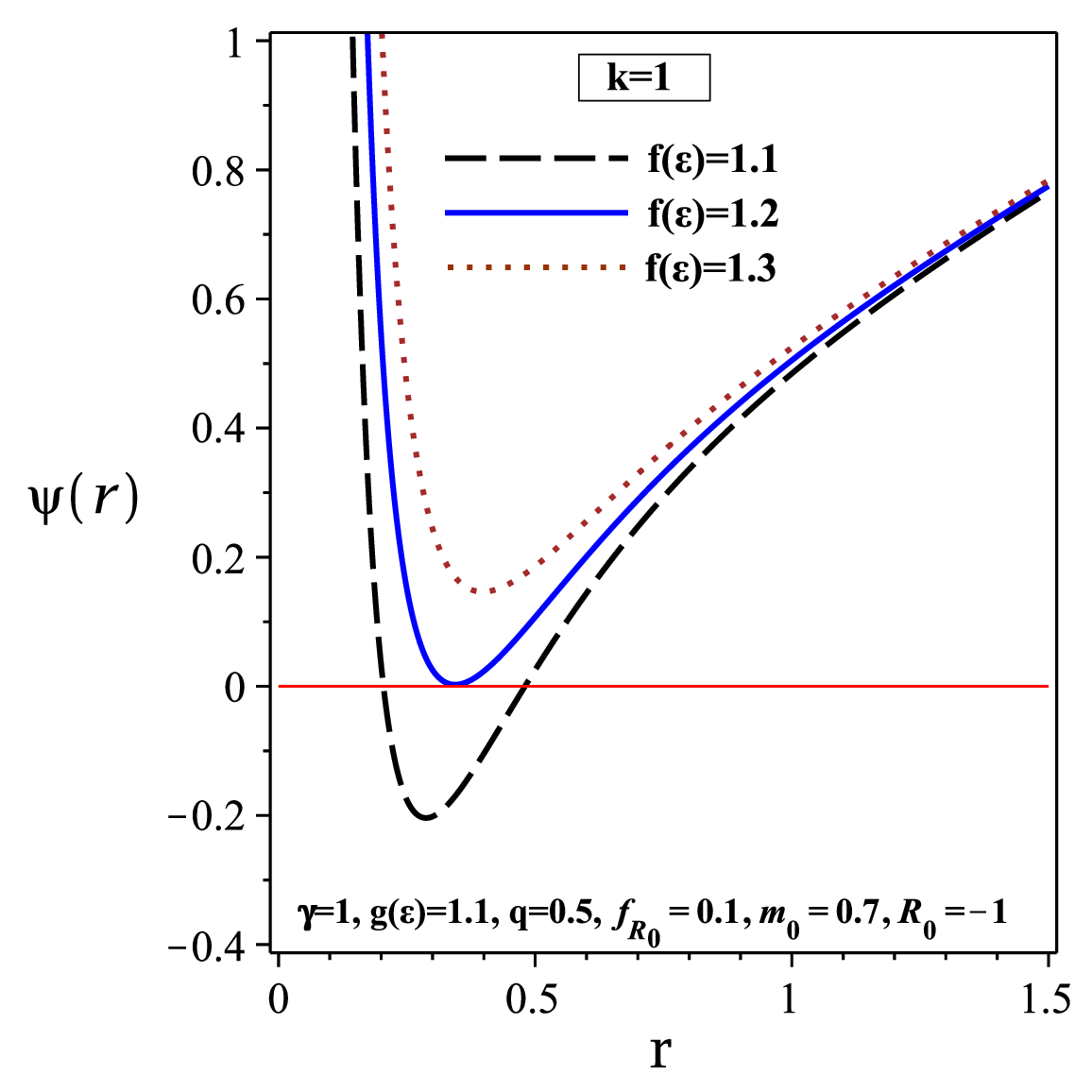} \includegraphics[width=0.3%
\textwidth]{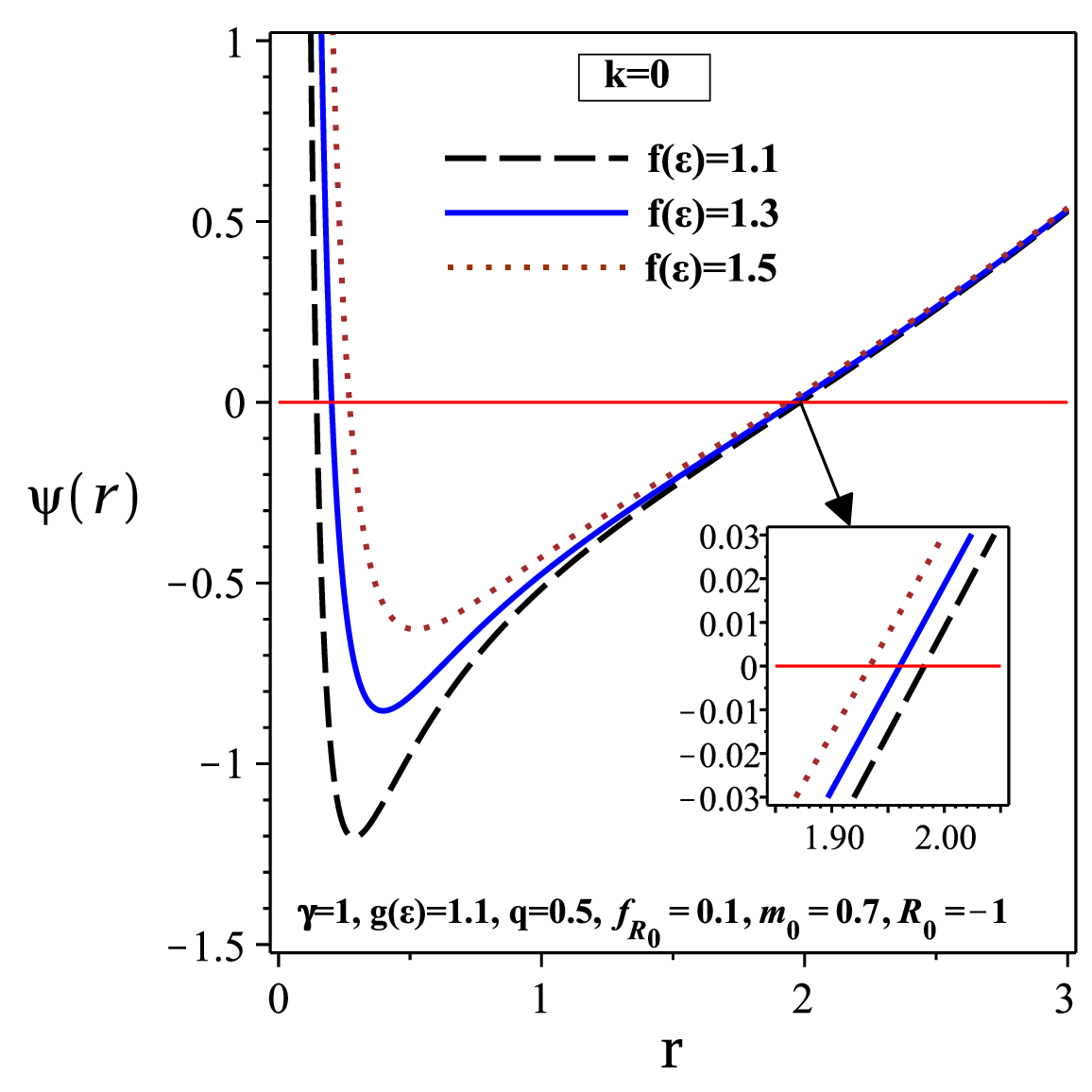} \includegraphics[width=0.3\textwidth]{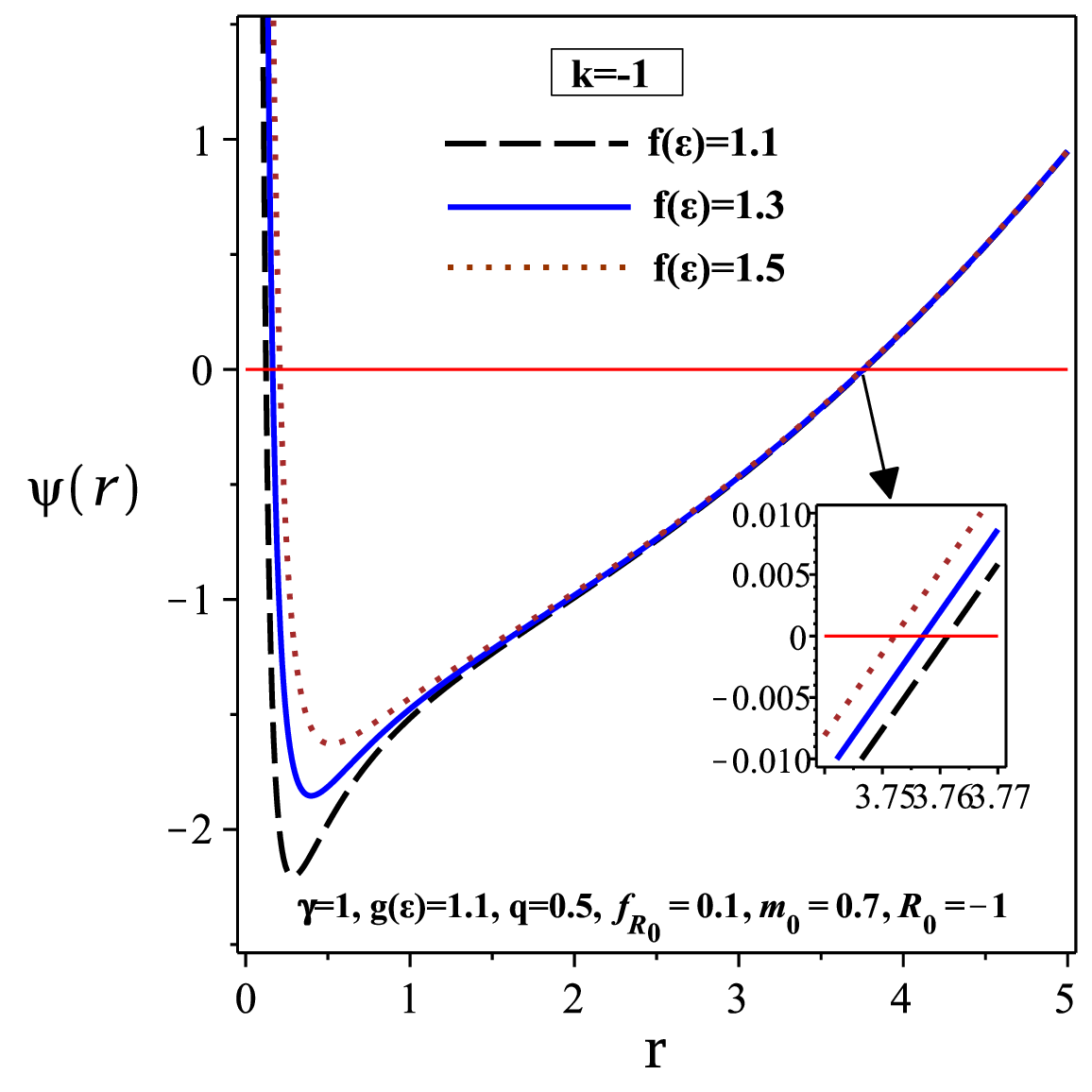} \newline
\caption{$\protect\psi(r)$ versus $r$ for different values of $k$, and $%
f\left( \protect\varepsilon \right)$.}
\label{Fig2}
\end{figure}

\begin{figure}[tbh]
\centering
\includegraphics[width=0.3\textwidth]{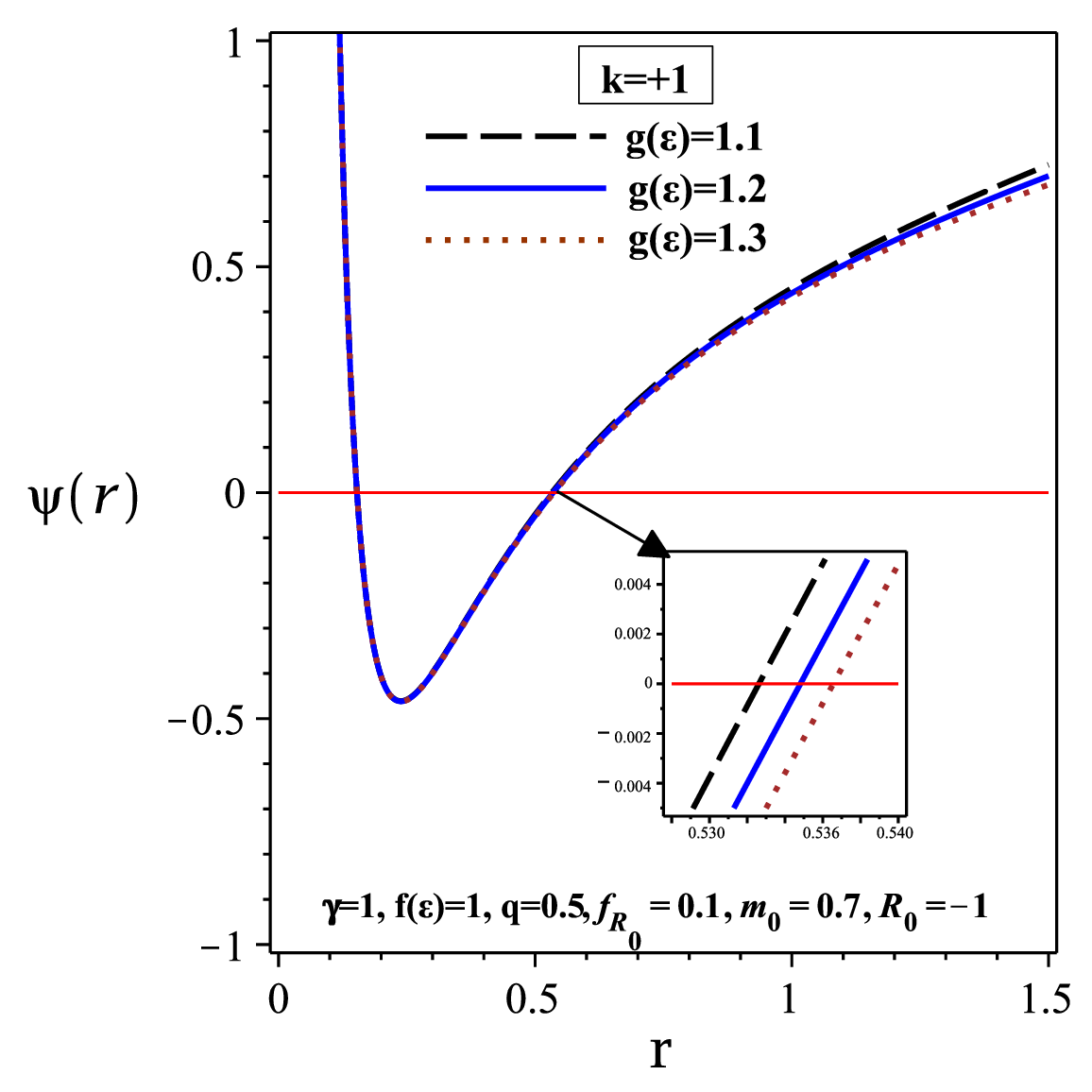} \includegraphics[width=0.3%
\textwidth]{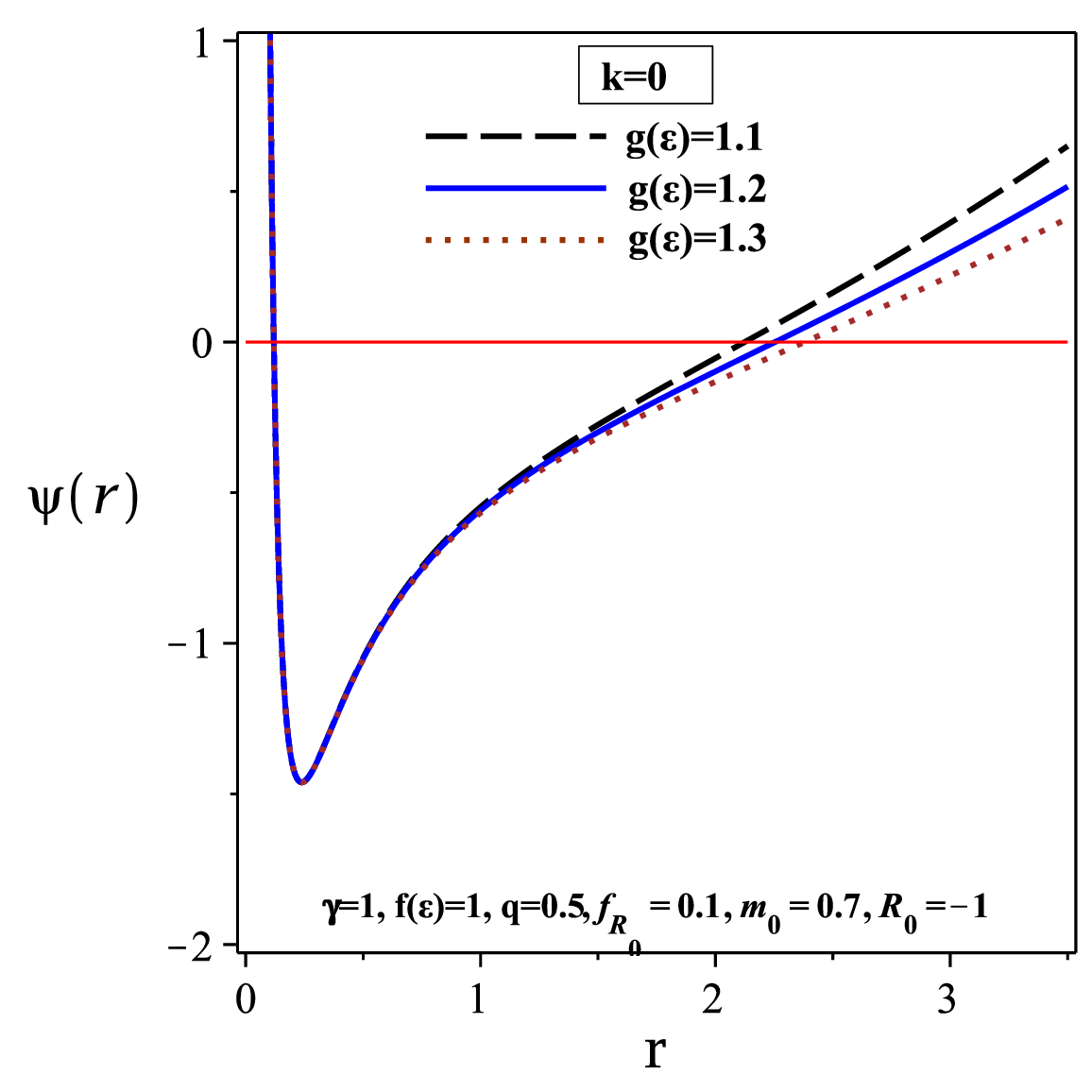} \includegraphics[width=0.3\textwidth]{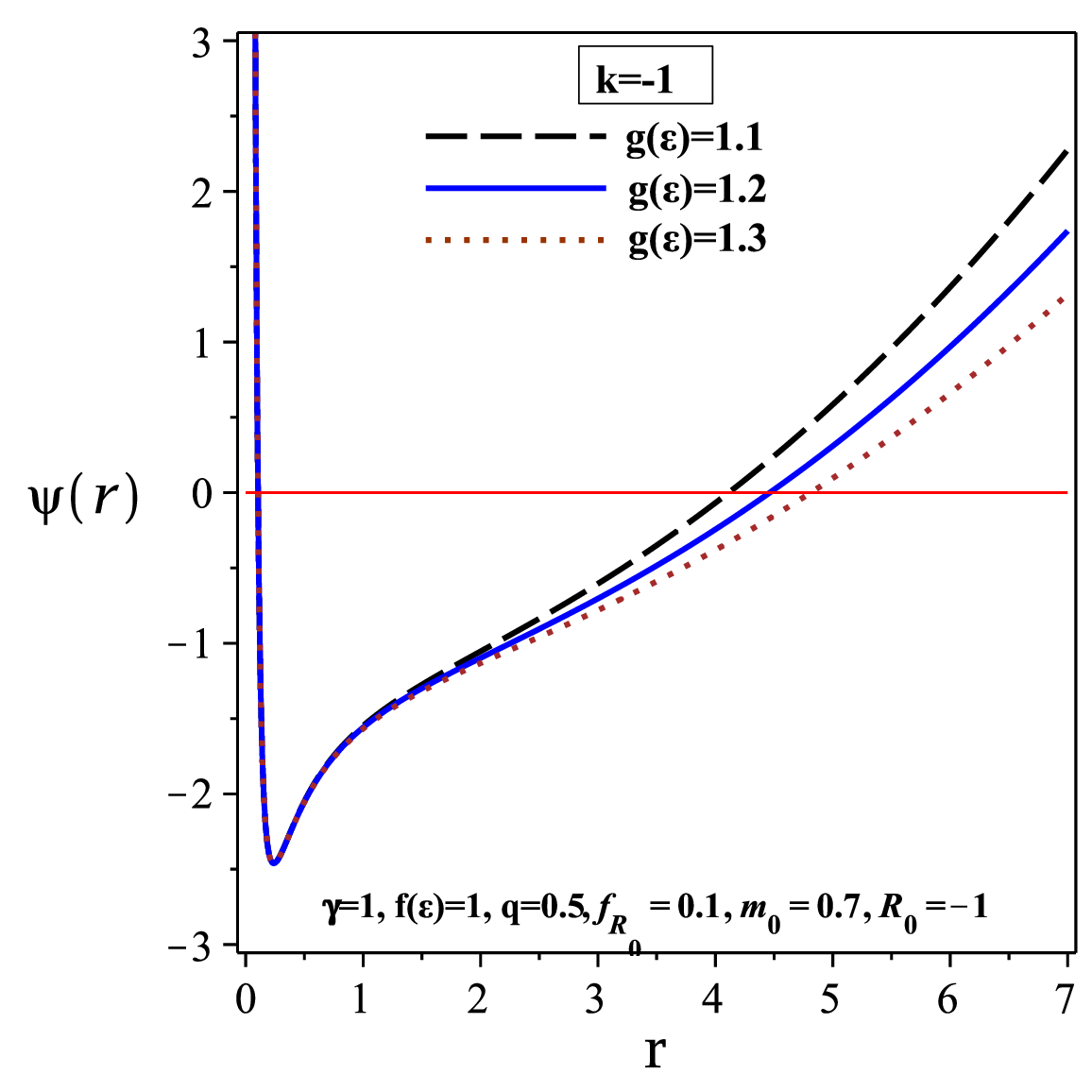} \newline
\caption{$\protect\psi(r)$ versus $r$ for different values of $k$, and $%
g\left( \protect\varepsilon \right)$.}
\label{Fig3}
\end{figure}

Here, we study the effects of the topological constant ($k$) and rainbow
functions $f\left( \varepsilon \right)$ and $g\left( \varepsilon \right)$ on
the obtained black holes. For this purpose, we plot Figs. \ref{Fig2} and \ref%
{Fig3}. Our findings indicate that the large black holes belong to case $%
k=-1 $. Moreover, as $f\left(\varepsilon\right)$ ($g\left(\varepsilon\right)$%
) increases, the radius of the event horizon decreases (increases). In
addition, by comparing the three panels in Figs. \ref{Fig2} and \ref{Fig3},
we can see that the radius of the event horizon becomes more sensitive to an
increase in $f\left(\varepsilon\right)$ ($g\left(\varepsilon\right)$) when $%
k=+1$ ($k=-1$).

\begin{figure}[tbh]
\centering
\includegraphics[width=0.3\textwidth]{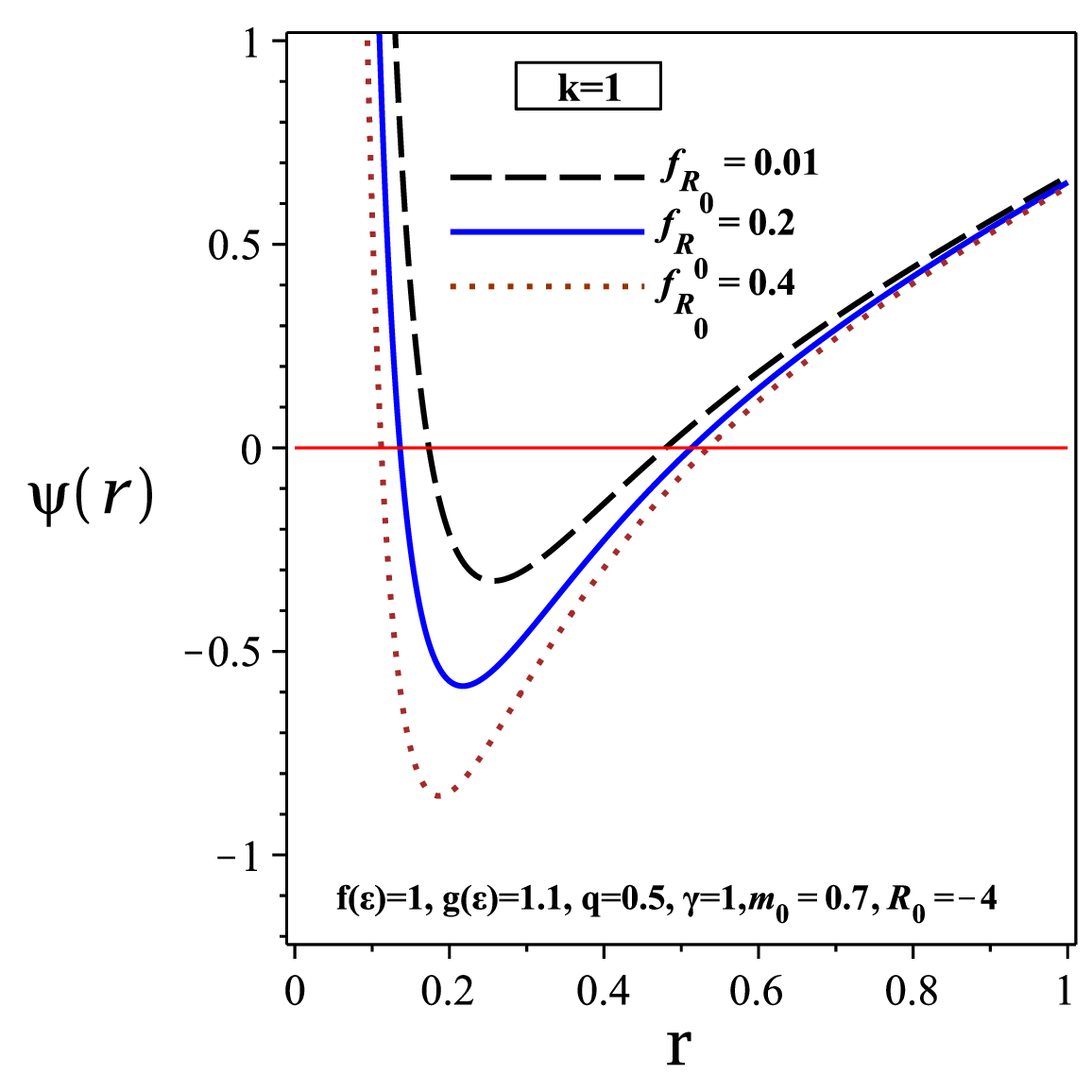} \includegraphics[width=0.3%
\textwidth]{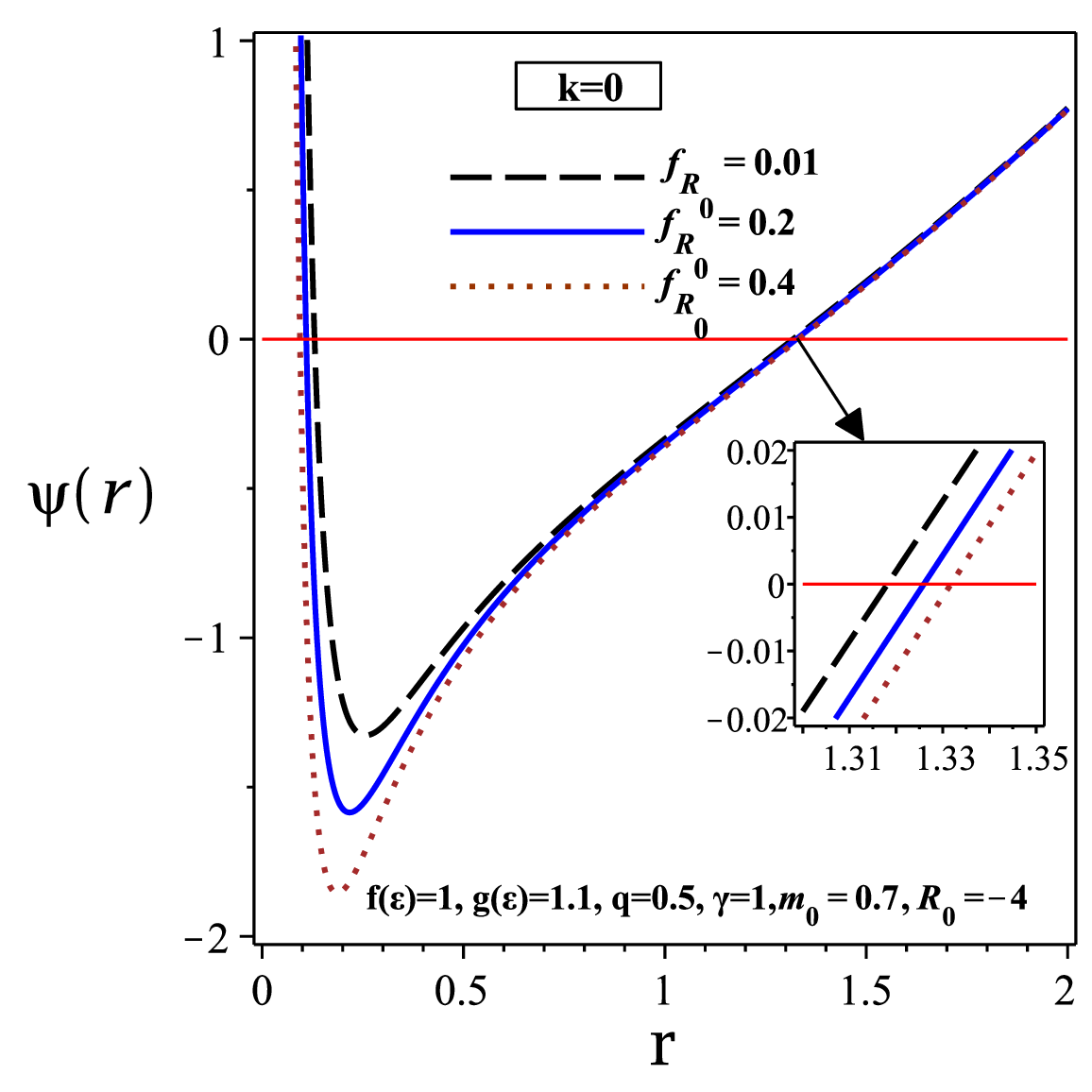} \includegraphics[width=0.3\textwidth]{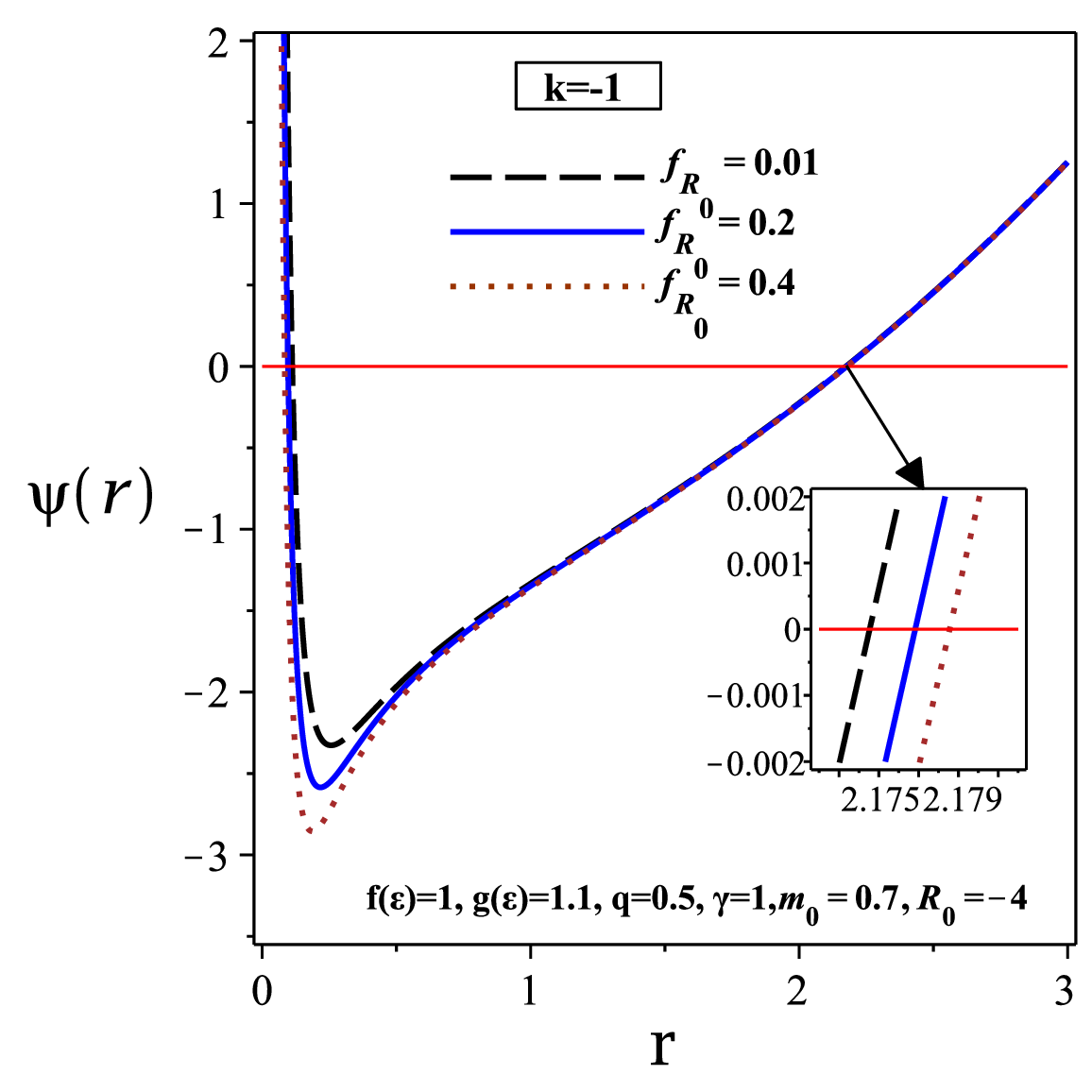} \newline
\caption{$\protect\psi(r)$ versus $r$ for different values of $k$, and $%
f_{R_{0}}$.}
\label{Fig4}
\end{figure}

We study the effect of $f_{R_{0}}$ (a parameter of $F(R)$ gravity) by
considering different topological constants in Fig. \ref{Fig4}. As can be
observed, the large black holes belong to the case $k=-1$, similar to the
previous cases. However, increasing the value of $f_{R_{0}}$ results in
larger black holes. In fact, the radius of the event horizon of the black
holes increases as the parameter of $F(R)$ gravity is increased. Moreover,
the radius of the black hole is more responsive in the case $k=+1$.

\begin{figure}[tbh]
\centering
\includegraphics[width=0.35\textwidth]{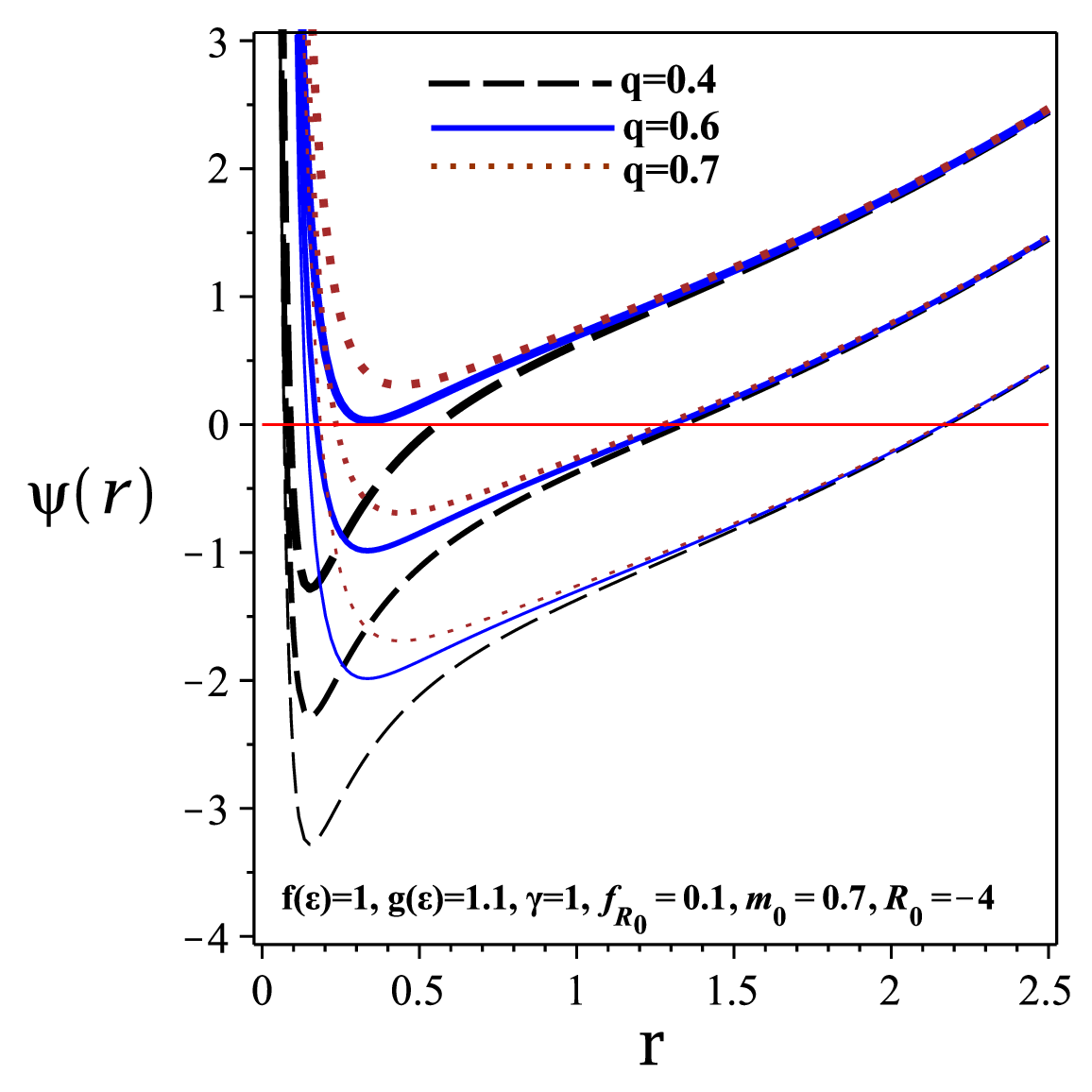} \includegraphics[width=0.35%
\textwidth]{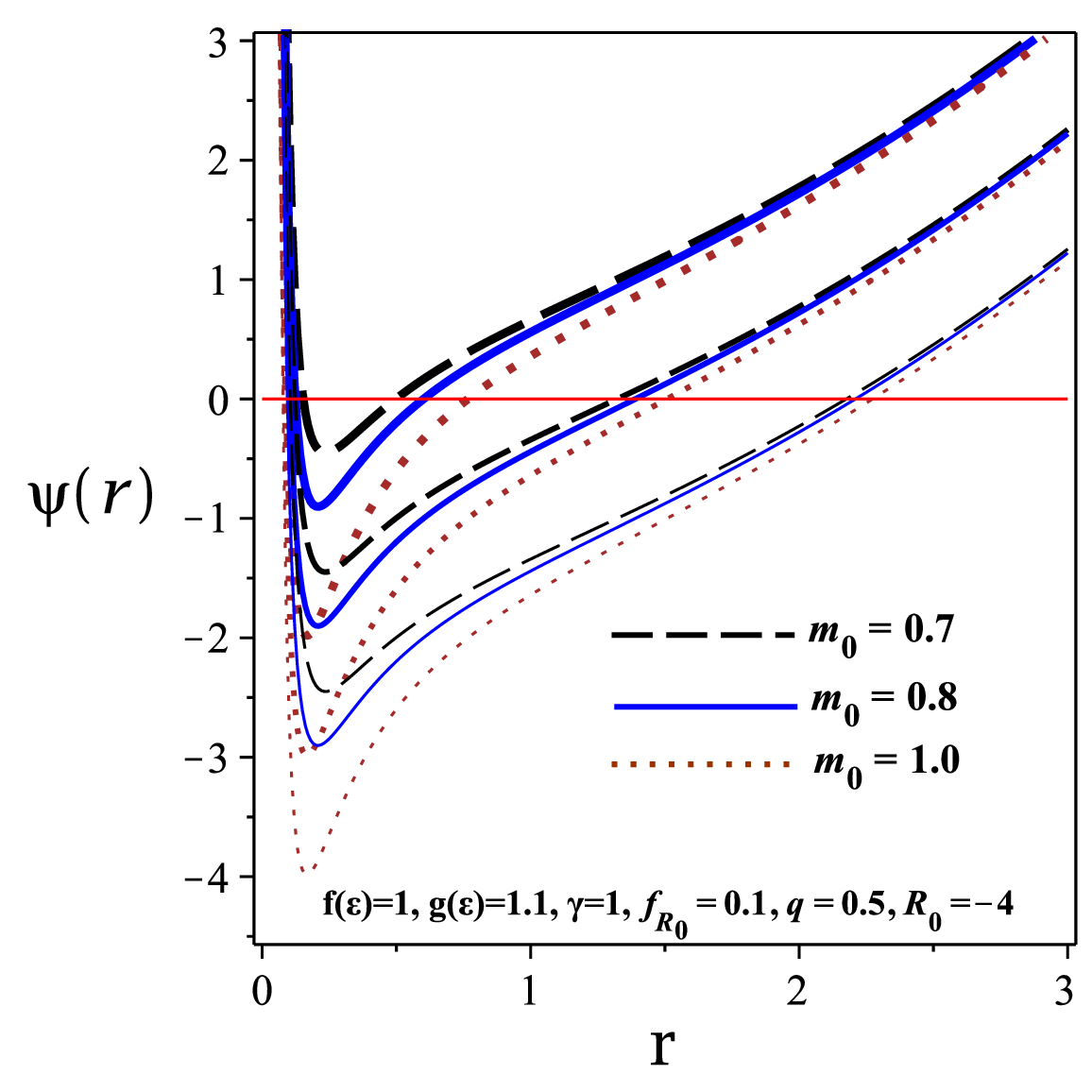} \newline
\caption{$\protect\psi(r)$ versus $r$ for different values of $k$, $q$ (left
panel), and $m_{0}$ (right panel). Three up panels are related to $k=+1$.
Three middle panels belong to $k=0$. Also, three down panels are plotted for 
$k=-1$.}
\label{Fig5}
\end{figure}

The effects of mass and electrical charge on the radius of the black hole is
plotted in Fig. \ref{Fig5}. Black holes with a large event horizon are
associated with the topological constant $k=-1$. As the mass and electrical
charge increase, the radius of the black hole's event horizon also
increases. In other words, larger black holes have greater masses and higher
electrical charges.

As a consequence of various parameters affecting the event horizon, larger
black holes are classified as black holes with a negative topological
constant ($k=-1$). These black holes possess greater mass, electric charge, $%
\gamma$, $f_{R_{0}}$, and $g\left(\varepsilon\right)$ but lower $%
f\left(\varepsilon\right)$. On the other hand, our findings indicate that
the radius of the event horizon is more sensitive for $k=+1$ compared to
other values of the topological constant (except when $g\left(\varepsilon%
\right)$ increases, see Fig. \ref{Fig3}).

\section{\textbf{Thermodynamics}}

\label{Thermodynamics}

Now, we are going to calculate the conserved and thermodynamic quantities of
the topological black hole solutions in $F(R)$ gravity's rainbow to check
the first law of thermodynamics.

For studying the thermodynamic properties of the obtained black hole
solutions, it is necessary to express the mass ($m_{0}$) in terms of the
radius of the event horizon $r_{+}$ and the charge $q$ as follows. Equating $%
g_{tt}=g(r)$\ to zero, we have 
\begin{equation}
m_{0}=kr_{+}-\frac{R_{0}r_{+}^{3}}{12g^{2}\left( \varepsilon \right) }+\frac{
q^{2}e^{-\gamma }f^{2}\left( \varepsilon \right) }{\left( 1+f_{R_{0}}\right)
r_{+}}.  \label{mm}
\end{equation}

Here, we want to obtain the Hawking temperature for these black holes. The
superficial gravity of a black hole is given by 
\begin{equation}
\kappa =\left. \frac{g_{tt}^{\prime }}{2\sqrt{-g_{tt}g_{rr}}}=\right\vert
_{r=r_{+}}=\left. \frac{\psi ^{\prime }(r)g\left( \varepsilon \right) }{%
2f\left( \varepsilon \right) }\right\vert _{r=r_{+}},  \label{k}
\end{equation}%
where $r_{+}$ is the radius of the events horizon. Considering the obtained
metric function (\ref{g(r)F(R)}), and by substituting the mass (\ref{mm})\
within the equation (\ref{k}), one can calculate the superficial gravity as 
\begin{equation}
\kappa =\frac{kg\left( \varepsilon \right) }{2f\left( \varepsilon \right)
r_{+}}-\frac{R_{0}r_{+}}{8g\left( \varepsilon \right) f\left( \varepsilon
\right) }-\frac{q^{2}e^{-\gamma }g\left( \varepsilon \right) f\left(
\varepsilon \right) }{2\left( 1+f_{R_{0}}\right) r_{+}^{3}},
\end{equation}%
and by using the Hawking temperature as $T=\frac{\kappa }{2\pi }$, we can
extract it in the following form 
\begin{equation}
T=\frac{kg\left( \varepsilon \right) }{4\pi f\left( \varepsilon \right) r_{+}%
}-\frac{R_{0}r_{+}}{16\pi g\left( \varepsilon \right) f\left( \varepsilon
\right) }-\frac{q^{2}e^{-\gamma }g\left( \varepsilon \right) f\left(
\varepsilon \right) }{4\pi \left( 1+f_{R_{0}}\right) r_{+}^{3}}.
\label{TemF(R)CPMI}
\end{equation}

Considering Eq. (\ref{TemF(R)CPMI}), we can study the behavior of the
temperature in both the limit $r_{+}\rightarrow 0$ (known as the high-energy
limit) and the limit $r_{+}\rightarrow \infty$ (known as the asymptotic
behavior). The high-energy limit of the temperature is given by 
\begin{equation}
\underset{r_{+}\rightarrow 0}{\lim }T\propto -\frac{q^{2}e^{-\gamma }g\left(
\varepsilon \right) f\left( \varepsilon \right) }{4\pi \left(
1+f_{R_{0}}\right) r_{+}^{3}},
\end{equation}
in which depends on the electrical charge, the parameter of the ModMax
theory, rainbow functions, and the parameter of $F(R)$ gravity. Notably, the
temperature is always negative in the high-energy limit. Therefore, the
small black holes in this theory of gravity are not physical objects.

We expand our study to investigate the asyptotical behavior of the
temperature, which leads to 
\begin{equation}
\underset{r_{+}\rightarrow \infty }{\lim }T\propto -\frac{R_{0}r_{+}}{16\pi
g\left( \varepsilon \right) f\left( \varepsilon \right) },
\end{equation}%
as one can see, the temperature of large black holes depends on the rainbow
functions ($g\left( \varepsilon \right) $ and $f\left( \varepsilon \right) $%
) and the constant scalar curvature ($R_{0}$). In order to have positive
large black holes, we must consider $R_{0}>0$.

When we consider $T=0$ (Eq. (\ref{TemF(R)CPMI})), we can find the
temperature's root (we show that by $r_{_{T=0}}$). Our analysis shows that
there is only one real root, which can be expressed as follows 
\begin{equation}
r_{_{T=0}}=\sqrt{\frac{2g\left( \varepsilon \right) \left( kg\left(
\varepsilon \right) \left( 1+f_{R_{0}}\right) -\sqrt{\left[ k^{2}g^{2}\left(
\varepsilon \right) \left( 1+f_{R_{0}}\right) -q^{2}e^{-\gamma }f^{2}\left(
\varepsilon \right) R_{0}\right] \left( 1+f_{R_{0}}\right) }\right) }{\left(
1+f_{R_{0}}\right) R_{0}}},
\end{equation}%
According to the above equation, the root temperature of the black holes
will become zero for a sufficiently large value of $\gamma $ (or in the
absence of an electrical charge).

As a result of our findings, in $F(R)$-ModMax gravity's rainbow, the
temperature of the large black holes can be positive and have only one root.
It is notable that, the temperature is negative before this root, and after
it, the temperature becomes positive.

The electric charge of black hole per unit volume, $\mathcal{V}$, can be
obtained by using the Gauss law as 
\begin{equation}
Q=\frac{\widetilde{Q}}{\mathcal{V}}=\frac{qf\left( \varepsilon \right) }{%
4\pi g\left( \varepsilon \right) }.  \label{Q}
\end{equation}

Considering $F_{\mu \nu }=\partial _{\mu }A_{\nu }-\partial _{\nu }A_{\mu }$
, one can find the nonzero component of the gauge potential in which is $%
A_{t}=-\int F_{tr}dr$, and therefore the electric potential at the event
horizon ($U$) with respect to the reference ($r\rightarrow \infty $) is
given by 
\begin{equation}
U=-\int_{r_{+}}^{+\infty }F_{tr}dr=\frac{qe^{-\gamma }}{r_{+}}.
\label{elcpoF(R)CPMI}
\end{equation}

In order to obtain the entropy of black holes in $F(R)=R+f(R)$\ theory, one
can use a modification of the area law which means the Noether charge method 
\cite{Cognola2005} 
\begin{equation}
S=\frac{A(1+f_{R_{0}})}{4},  \label{SFR}
\end{equation}%
where $A$\ is the horizon area and is defined 
\begin{equation}
A=\left. \int_{0}^{2\pi }\int_{0}^{\pi }\sqrt{g_{\theta \theta }g_{\varphi
\varphi }}\right\vert _{r=r_{+}}=\left. \frac{r^{2}}{g^{2}\left( \varepsilon
\right) }\right\vert _{r=r_{+}}=\frac{r_{+}^{2}}{g^{2}\left( \varepsilon
\right) },  \label{A}
\end{equation}%
so, the entropy of topological phantom AdS black holes per unit volume, $%
\mathcal{V}$, in $F(R)$ gravity is given by replacing the horizon area (\ref%
{A}) within Eq. (\ref{SFR}) as

\begin{equation}
S=\frac{\widetilde{S}}{\mathcal{V}}=\frac{(1+f_{R_{0}})r_{+}^{2}}{%
4g^{2}\left( \varepsilon \right) },  \label{S}
\end{equation}%
which indicates that the area law does not hold for the black hole solutions
in $R+f(R)$ gravity.

Using Ashtekar-Magnon-Das (AMD) approach \cite{AMDI,AMDII}, we find the
total mass of these black holes per unit volume, $\mathcal{V}$, in $F(R)$
gravity as 
\begin{equation}
M=\frac{\widetilde{M}}{\mathcal{V}}=\frac{m_{0}\left( 1+f_{R_{0}}\right) }{%
8\pi g\left( \varepsilon \right) f\left( \varepsilon \right) },
\label{AMDMass}
\end{equation}%
where substituting the mass (\ref{mm}) within the equation (\ref{AMDMass}),
yields 
\begin{equation}
M=\frac{\left( 1+f_{R_{0}}\right) r_{+}}{8\pi g\left( \varepsilon \right)
f\left( \varepsilon \right) }\left( k-\frac{R_{0}r_{+}^{2}}{12g^{2}\left(
\varepsilon \right) }\right) +\frac{q^{2}e^{-\gamma }f\left( \varepsilon
\right) }{8\pi g\left( \varepsilon \right) r_{+}}.  \label{MM}
\end{equation}

Here, we study the high-energy limit and asymptotical behavior of the mass.
The high-energy limit of the mass is 
\begin{equation}
\underset{r_{+}\rightarrow 0}{\lim }M\propto \frac{q^{2}e^{-\gamma }f\left(
\varepsilon \right) }{8\pi g\left( \varepsilon \right) r_{+}},
\end{equation}%
where indicates that the high-energy limit depends on several factors,
including the electrical charge ($q$), the ModMax theory ($\gamma $), the
rainbow functions ($f\left( \varepsilon \right) $ and $g\left( \varepsilon
\right) $), and the parameter of $F(R)$ gravity. Furthermore, it is
important to note that the mass is always positive in this limit.

The asymptotical behavior is given by 
\begin{equation}
\underset{r_{+}\rightarrow \infty }{\lim }M\propto -\frac{\left(
1+f_{R_{0}}\right) R_{0}r_{+}^{3}}{96\pi g^{3}\left( \varepsilon \right)
f\left( \varepsilon \right) },
\end{equation}%
where it is always positive because $R_{0}<0$. As mentioned earlier in
relation to temperature, in order to have a positive temperature, we need to
consider $R_{0}<0$. We now apply this condition to analyze the asymptotical
behavior of mass. Consequently, the mass will always be positive as $%
r_{+}\rightarrow \infty $.

On the other hand, our analysis reveals an interesting behavior for middle
black holes when $k=-1$. Specifically, the mass of middle black holes
depends on the value of the topological constant. That is, $%
M_{middle}\propto \frac{k\left( 1+f_{R_{0}}\right) r_{+}}{8\pi g\left(
\varepsilon \right) f\left( \varepsilon \right) }$, and this relationship
varies for different values of $k$, i.e., 
\begin{equation}
M_{middle}\propto \frac{k\left( 1+f_{R_{0}}\right) r_{+}}{8\pi g\left(
\varepsilon \right) f\left( \varepsilon \right) }=\left\{ 
\begin{array}{cc}
\frac{\left( 1+f_{R_{0}}\right) r_{+}}{8\pi g\left( \varepsilon \right)
f\left( \varepsilon \right) } & k=+1 \\ 
&  \\ 
0 & k=0 \\ 
&  \\ 
\frac{-\left( 1+f_{R_{0}}\right) r_{+}}{8\pi g\left( \varepsilon \right)
f\left( \varepsilon \right) } & k=-1%
\end{array}%
\right. ,
\end{equation}%
where indicates that the total mass is always positive for $k=+1$ and $k=0$.
However, for $k=-1$, small and large black holes have positive mass, while
the middle black holes encounter negative mass. This implies that there are
two roots for $k=-1$, but no roots for $k=+1$ or $k=0$. The roots of the
mass ($r_{1,2_{M=0}}$) for $k=-1$ are given by 
\begin{equation}
r_{1,2_{M=0}}=\sqrt{\frac{6g\left( \varepsilon \right) \left( -g\left(
\varepsilon \right) \left( 1+f_{R_{0}}\right) \pm \sqrt{\left[ g^{2}\left(
\varepsilon \right) \left( 1+f_{R_{0}}\right) +\frac{q^{2}e^{-\gamma
}f^{2}\left( \varepsilon \right) R_{0}}{3}\right] \left( 1+f_{R_{0}}\right) }%
\right) }{\left( 1+f_{R_{0}}\right) R_{0}}}.
\end{equation}%
where $r_{1_{M=0}}$ and $r_{2_{M=0}}$ refer to the smaller and larger roots
of the mass, respectively, our findings reveal that the mass of a black hole
is positive in the ranges $r_{+}<r_{1_{M=0}}$ and $r_{+}>r_{2_{M=0}}$.
Additionally, it is negative between $r_{1_{M=0}}$ and $r_{2_{M=0}}$ (i.e., $%
r_{1_{M=0}}<r_{+}<r_{2_{M=0}}$).

It is straightforward to show that the conserved and thermodynamics
quantities satisfy the first law of thermodynamics 
\begin{equation}
dM=TdS+UdQ,
\end{equation}%
where $T=\left( \frac{\partial M}{\partial S}\right) _{Q}$, and $U=\left( 
\frac{\partial M}{\partial Q}\right) _{S}$, and they are in agreement with
those of calculated in Eqs. (\ref{TemF(R)CPMI}) and (\ref{elcpoF(R)CPMI}),
respectively.


\section{\textbf{Thermodynamic Topology and Duan's $\protect\phi$ mapping
theory}}

\label{tt} The main concept related to defects is the topological charge. In
order to analyze the thermodynamic topology of a black hole, we compute the
topological charge and use it to identify the topological classes. The
specific method we employ to calculate the topological charge is known as
Duan's $\phi $ mapping technique. The mathematical steps required for Duan's 
$\phi $ mapping technique are outlined in the following paragraph. From the
expression for the off-shell free energy of a black hole a vector field is
constructed as follows \cite{29} 
\begin{equation}
\phi =\left( \phi ^{r},\phi ^{\Theta }\right) =\left( \frac{\partial 
\mathcal{F}}{\partial r_{+}},-\cot \Theta ~\csc \Theta \right) ,
\label{phi1}
\end{equation}%
in the $\phi ^{\Theta }$ component, the trigonometric function is chosen so
that one zero point of the vector field can always be found at $\theta =%
\frac{\pi }{2}$. The other zero point can also be found by simply solving
the equation, $\phi ^{r}=0$, which always results in $\tau =\frac{1}{T}$.
The basic topological property associated with the zero point or topological
defect of a field is its winding number or topological charge. In this work,
we use Duan's $\phi $ mapping technique \cite{g1,g2} to calculate the
winding number. To find the topological charge we first determine the unit
vector $n$ of the field in Eq. (\ref{phi1}), which are

\begin{eqnarray}
n^{1} &=&\frac{\phi ^{r}}{\sqrt{(\phi ^{r})^{2}+(\phi ^{\Theta })^{2}}}, 
\notag \\
&&  \label{ns} \\
n^{2} &=&\frac{\phi ^{\Theta }}{\sqrt{(\phi ^{r})^{2}+(\phi ^{\Theta })^{2}}}%
.  \notag
\end{eqnarray}

For the vector field, a topological current can be constructed in the
coordinate space $x^\nu=\{t,r_+,\theta\}$ as follows \cite{g1,g2} 
\begin{equation}
j^\mu=\frac{1}{2\pi}\epsilon^{\mu \nu \rho}\epsilon_{ab}\partial_\nu
n^a\partial_\rho n^b,  \label{current}
\end{equation}
where $\partial_\nu=\frac{\partial}{\partial x^\nu}$ and $\mu,\nu,\rho=0, 1,
2$. The fundamental conditions that have to be fulfilled by the normalized
vector $n^a$ are \cite{g1,g2} 
\begin{equation}
n^an^a=1 \quad \text{and} \quad n^a\partial_\nu n^a=0.
\end{equation}

The current given in Eq. (\ref{current}), is a conserved quantity, which can
be verified by applying the current conservation law 
\begin{equation}
\partial_\mu j^\mu=0.
\end{equation}

It can be proved that the current $j^{\mu }$ is only non-zero at the zero
points of the vector field by using the following equation \cite{29,g1,g2} 
\begin{equation}
j^{\mu }=\delta ^{2}(\phi )J^{\mu }\left( \frac{\phi }{x}\right) ,
\end{equation}%
where we use the following properties of Jacobi tensor 
\begin{equation}
\epsilon ^{ab}J^{\mu }\left( \frac{\phi }{x}\right) =\epsilon ^{\mu \nu \rho
}\partial _{\nu }\phi ^{a}\partial _{\rho }\phi ^{b},
\end{equation}%
and the two-dimensional Laplacian Green function 
\begin{equation}
\Delta _{\phi ^{a}}\ln ||\phi ||=2\pi \delta ^{2}(\phi ).
\end{equation}

Again, the topological charge $W$ is related to the $0^{th}$ component of
the current density of the topological current through the following
relation 
\begin{equation}
W=\int_{\Sigma }j^{0}d^{2}x=\sum_{i=1}^{N}\beta _{i}\eta
_{i}=\sum_{i=1}^{N}w_{i},
\end{equation}%
where $w_{i}$ is the winding number around the zero point. Also, $\beta {i}$
and $\eta {i}$ are the Hopf index and the Brouwer degree, respectively. The
detailed derivation of the above formula can be referred to \cite{g1,g2} or 
\cite{29}.

Hence, the topological charge is also nonzero only at the zero points of the
vector field. To find the exact zero point where the topological charge is
to be calculated, we plot the unit vector field $n$ and find out the zero
point at which it diverges. The zero point always turns out to be $(\frac{1}{%
T},\frac{\pi }{2}).$ Next, a contour is chosen around each zero point and is
parametrized as 
\begin{equation}
\begin{cases}
r_{+}=r_{1}cos\nu +r_{0}, \\ 
\\ 
\theta =r_{2}sin\nu +\frac{\pi }{2},%
\end{cases}%
\end{equation}%
where $\nu \in (0,2\pi )$. In addition, $r_{1}$ and $r_{2}$ are the
parameters that determine the size of the contour to be drawn. Also, $r_{0}$
is the point around which the contour is drawn. $r_{1}, r_{2}$ and $r_{0}$
are chosen in such a way that the contour $C$ encloses the defect or zero
point of the vector field $n$. After that, the deflection of the vector
field $n$ is found along the contour $C$ as \cite{28,29} 
\begin{equation}
\Omega (\nu )=\int_{0}^{\nu }\epsilon _{12}n^{1}\partial _{\nu }n^{2}d\nu ,
\end{equation}%
this is followed by the calculation of the winding number $w_{i}$ around the 
$i^{th}$ zero point of the vector field, as follows 
\begin{equation}
w=\frac{\Omega (2\pi )}{2\pi }.
\end{equation}

Finally, the topological charge $W$ can be determined by summing the winding
numbers calculated along each contour around the zero points, i.e. 
\begin{equation}
W=\sum_{i}w_{i}.
\end{equation}

It is worth noting that when the parameter region does not have any zero
points, the total topological charge is set to zero.

In the next two subsections, we will examine the thermodynamic topology of
topological black holes in $F(R)$-ModMax gravity's rainbow under two
different ensembles: the fixed charge $q$ ensemble and the fixed potential $%
\phi$ ensemble. For each ensemble, we will determine the topological charge
and analyze the impact of the ensemble choice and its corresponding
thermodynamic parameter on the topological charge.

\subsection{\textbf{For elliptic (}$k=1$\textbf{) curvature hypersurface}}

\subsubsection{\textbf{Fixed charge (}$q$\textbf{) ensemble}}

For this ensemble, we calculate the off-shell free energy from equations (%
\ref{S}) and (\ref{MM}), which leads to 
\begin{equation}
\mathcal{F}=M-\frac{S}{\tau }=\frac{\frac{q^{2}e^{-\gamma }g(\varepsilon
)f^{2}(\varepsilon )\tau }{(1+f_{R_{0}})}-2\pi f(\varepsilon
)r_{+}^{3}+g(\varepsilon )\left( k-\frac{r_{+}^{2}R_{0}}{12g^{2}(\varepsilon
)}\right) \tau r_{+}^{2}}{\frac{8\pi g^{2}(\varepsilon )f(\varepsilon )\tau
r_{+}}{(1+f_{R_{0}})}}.
\end{equation}

The components of the vector $\phi $ are found to be 
\begin{eqnarray}
\phi ^{r} &=&\frac{\left( g(\varepsilon )\left( g(\varepsilon )k\tau -\pi
f(\varepsilon )r_{+}\right) -\frac{r_{+}^{2}R_{0}}{4}\tau \right) r_{+}^{2}-%
\frac{q^{2}e^{-\gamma }g^{2}(\varepsilon )f^{2}(\varepsilon )\tau }{%
(1+f_{R_{0}})}}{\frac{8\pi g^{3}(\varepsilon )f(\varepsilon )\tau r_{+}^{2}}{%
(1+f_{R_{0}})}}, \\
&&  \notag \\
\phi ^{\Theta } &=&-\cot \Theta ~\csc \Theta .
\end{eqnarray}

The unit vectors $\left( n^{1},n^{2}\right) $ are computed using the
equation (\ref{ns}). Now we calculate the zero points of the $\phi ^{r}$
component by solving the following equation ($\phi ^{r}=0$) and find an
expression for $\tau $ in the following form 
\begin{equation}
\tau =\frac{4\pi (1+f_{R_{0}})g(\varepsilon )f\left( \varepsilon \right)
r_{+}^{3}}{kg^{2}(\varepsilon )\left( 1+f_{R_{0}}\right)
r_{+}^{2}-q^{2}e^{-\gamma }g^{2}(\varepsilon )f^{2}(\varepsilon )-\frac{%
R_{0}\left( 1+f_{R_{0}}\right) r_{+}^{4}}{4}}.
\end{equation}

\begin{figure}[h]
\centering
\begin{subfigure}{0.3\textwidth}
\includegraphics[width=5cm,height=5cm]{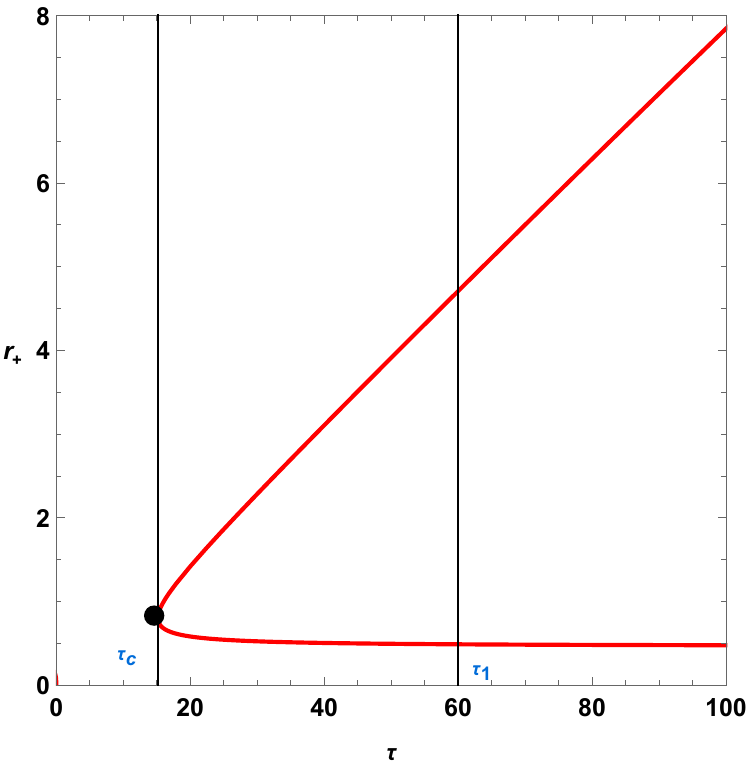}
\caption{}
\label{t1a}
\end{subfigure}
\begin{subfigure}{0.37\textwidth}
\includegraphics[height=5cm,width=6cm]{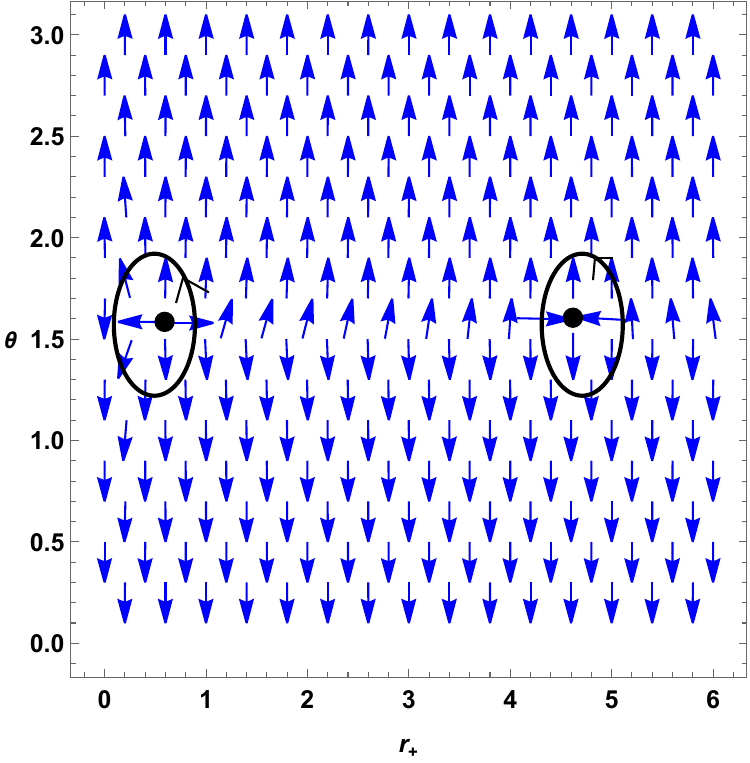}
\caption{}
\label{t1b}
\end{subfigure}
\begin{subfigure}{0.3\textwidth}
\includegraphics[width=5cm,height=5cm]{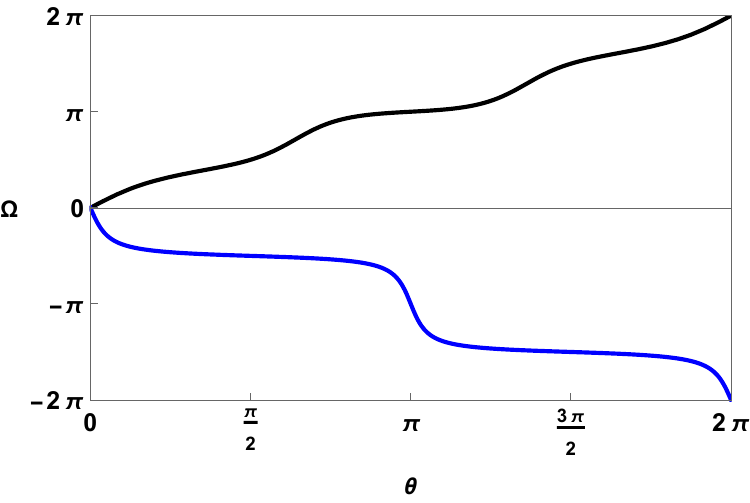}
\caption{}
\label{t1c}
\end{subfigure}
\caption{Topological charge of topological black holes $(k=1)$ in fixed
charge ensemble when positive values of $R_{0}$ is considered.}
\label{c1}
\end{figure}

Next, we plot the horizon radius $r_+$ against $\tau$ in Fig. \ref{t1a} for $%
f(\varepsilon)=g(\varepsilon)=1.2$, $q=0.5$, $R_{0}=0.001$, $\gamma =0.5$
and $k=1$. We observe two branches of black holes. In Fig. \ref{t1b}, a
vector plot is shown for the components $\phi ^{r}$ and $\phi ^{\theta }$,
taking $\tau =\tau{1}=60$. The zero points of the vector field are observed
at $r_{+}=0.49089$ and $r_{+}=4.70972$. From Fig. \ref{t1c}, it can be seen
that the winding number or topological charge corresponding to $%
r_{+}=0.49089 $ is $+1$, and the topological charge corresponding to $%
r_{+}=4.70972$ is $-1 $, represented by the solid lines in black and blue
colors, respectively, in Fig. \ref{t1c}. Hence, the total topological charge
is $1-1=0$. Additionally, an annihilation point is observed at $(\tau{c}%
,r_{+})=(15.1828, 0.8052)$, represented by the black dot in Fig. \ref{t1a}.

\begin{figure}[h]
\centering
\begin{subfigure}{0.3\textwidth}
\includegraphics[width=\linewidth]{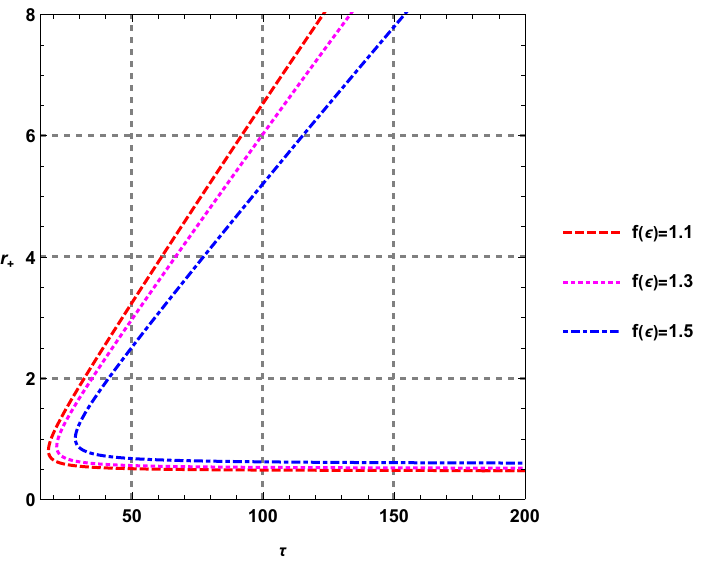}
\caption{}
\label{t2a}
\end{subfigure}
\begin{subfigure}{0.3\textwidth}
\includegraphics[width=\linewidth]{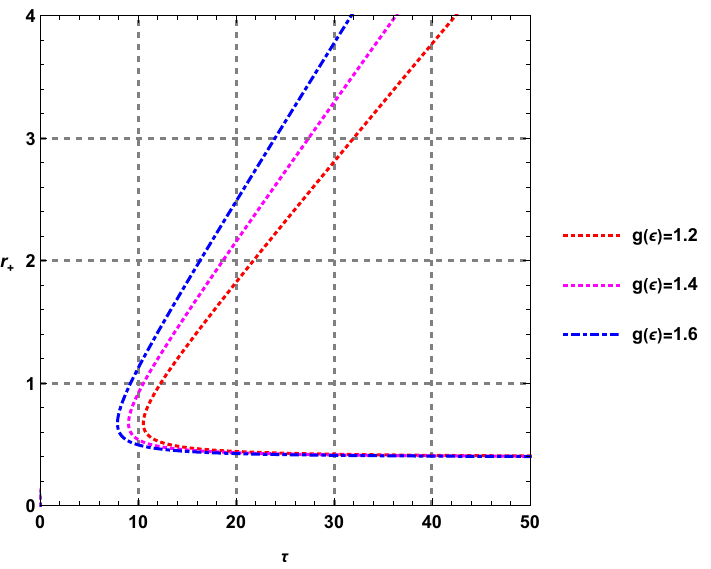}
\caption{}
\label{t2b}
\end{subfigure} \hspace{0.6cm} 
\begin{subfigure}{0.3\textwidth}
\includegraphics[width=\linewidth]{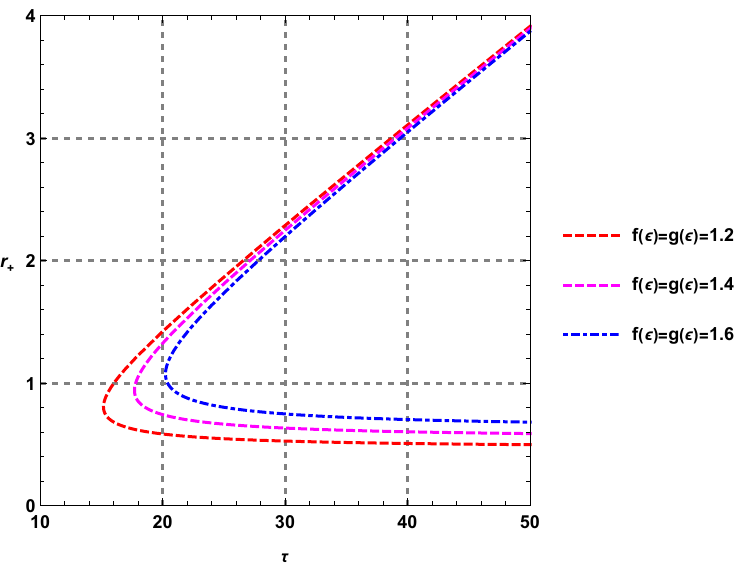}
\caption{}
\label{t2c}
\end{subfigure}
\begin{subfigure}{0.3\textwidth}
\includegraphics[width=\linewidth]{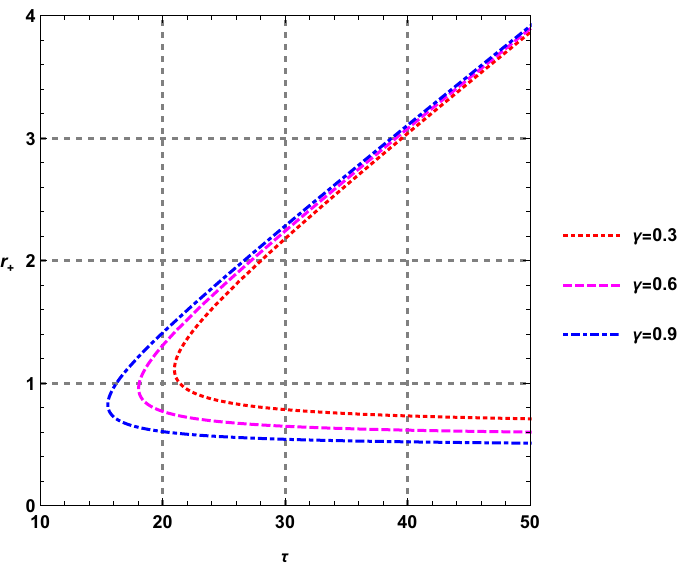}
\caption{}
\label{t2d}
\end{subfigure}
\begin{subfigure}{0.3\textwidth}
\includegraphics[width=\linewidth]{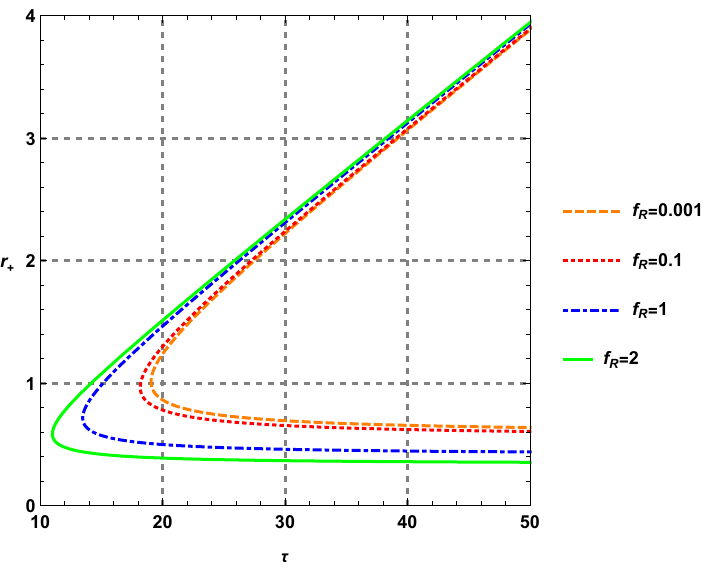}
\caption{}
\label{t2e}
\end{subfigure}
\caption{$\protect\tau $ vs $r_{+}$ plots for topological black holes $(k=1)$
in fixed charge ensemble when positive values of $R_{0}$ is considered. }
\label{c2}
\end{figure}

The topological charge remains invariant when varying the rainbow function,
represented by $g(\varepsilon)$, $f(\varepsilon)$, $\gamma$, and $f_{R_{0}}$
while keeping $q=0.5$, $R_{0}=0.001$, and $k=1$ fixed. In Figure. \ref{t2a},
the variation of $f(\varepsilon)$ is shown with $g(\varepsilon)=1$, $%
\gamma=0.5$, and $f_{R_{0}}=0.01$ fixed. In Figure. \ref{t2b}, the variation
of $g(\varepsilon)$ is shown with $f(\varepsilon)=1$, $\gamma=0.5$, and $%
f_{R_{0}}=0.01$ fixed. Figure. \ref{t2c} demonstrates the variation of both $%
f(\varepsilon)$ and $g(\varepsilon)$, while keeping $\gamma=0.5$ and $%
f_{R_{0}}=0.01$ fixed. In Figure. \ref{t2d}, the variation of $\gamma$ is
shown with $f(\varepsilon)=1.1,$ $g(\varepsilon)=1.1$, and $f_{R_{0}}=0.01$
fixed. Lastly, Figure. \ref{t2e} illustrates the variation of $f_{R_{0}}$
with $f(\varepsilon)=1.5$, $g(\varepsilon)=1.5$, and $\gamma=0.5$ fixed.
Based on Figure. \ref{c2}, it is evident that the topological charge is
consistently $0$ across all cases.

\begin{figure}[h]
\centering
\includegraphics[width=0.4\linewidth]{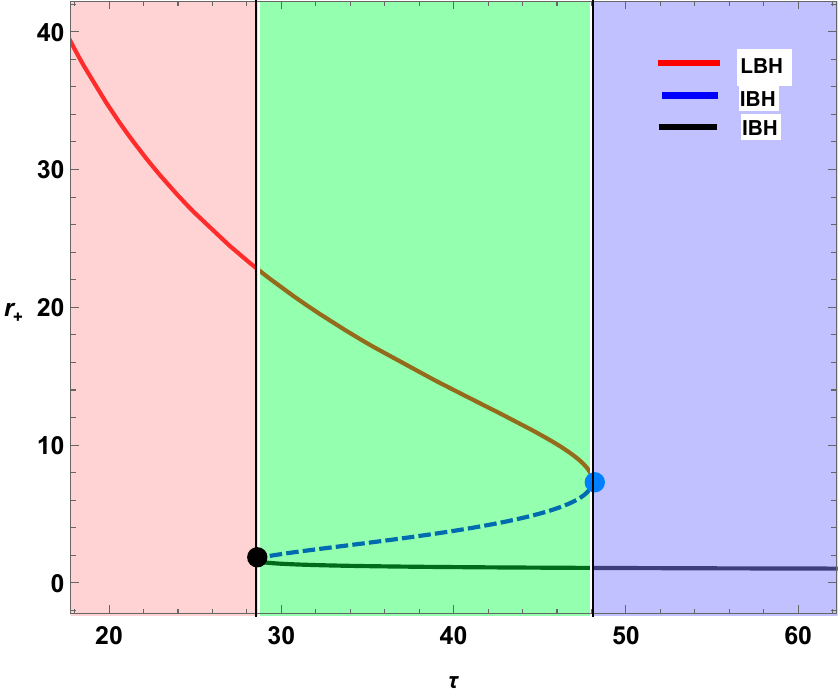} \label{c3a}
\caption{$\protect\tau $ vs $r_{+}$ plots for topological black holes $(k=1)$
in fixed charge ensemble for negative values of $R_{0}$.}
\label{c33}
\end{figure}

\begin{figure}[h]
\centering
\begin{subfigure}{0.35\textwidth}
\includegraphics[width=\linewidth]{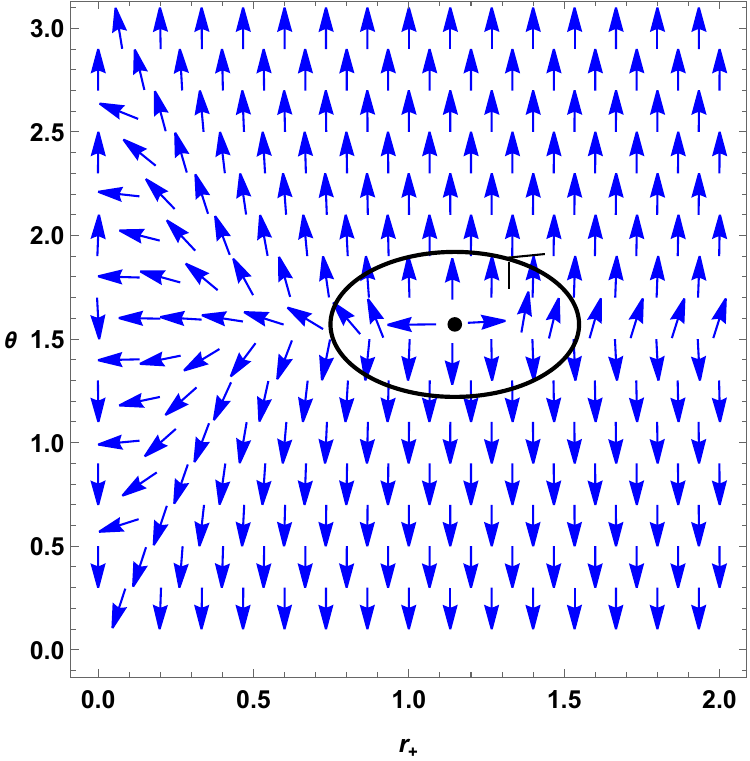}
\caption{}
\label{c3b}
\end{subfigure} 
\begin{subfigure}{0.35\textwidth}
\includegraphics[width=\linewidth]{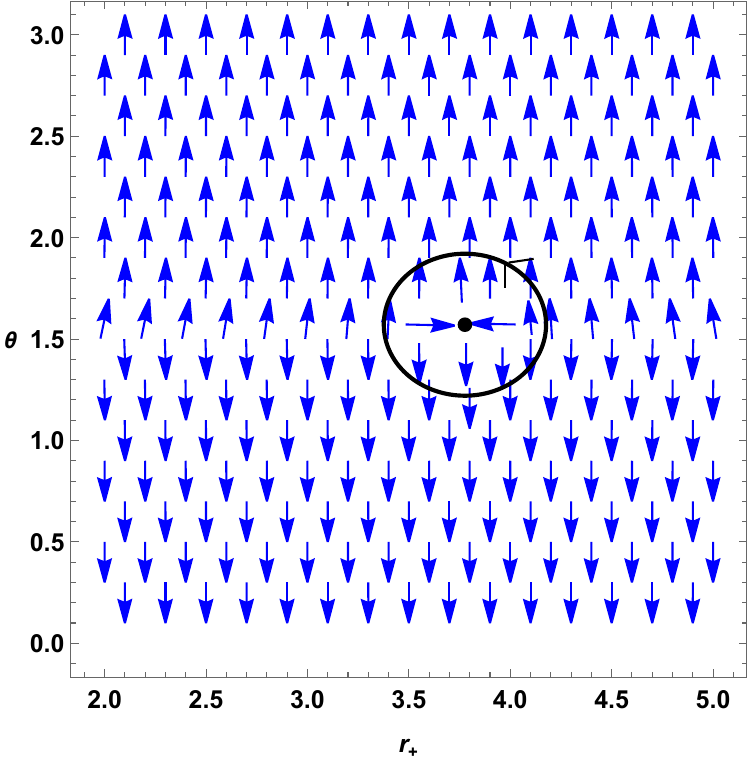}
\caption{}
\label{c3c}
\end{subfigure}
\begin{subfigure}{0.35\textwidth}
\includegraphics[width=\linewidth]{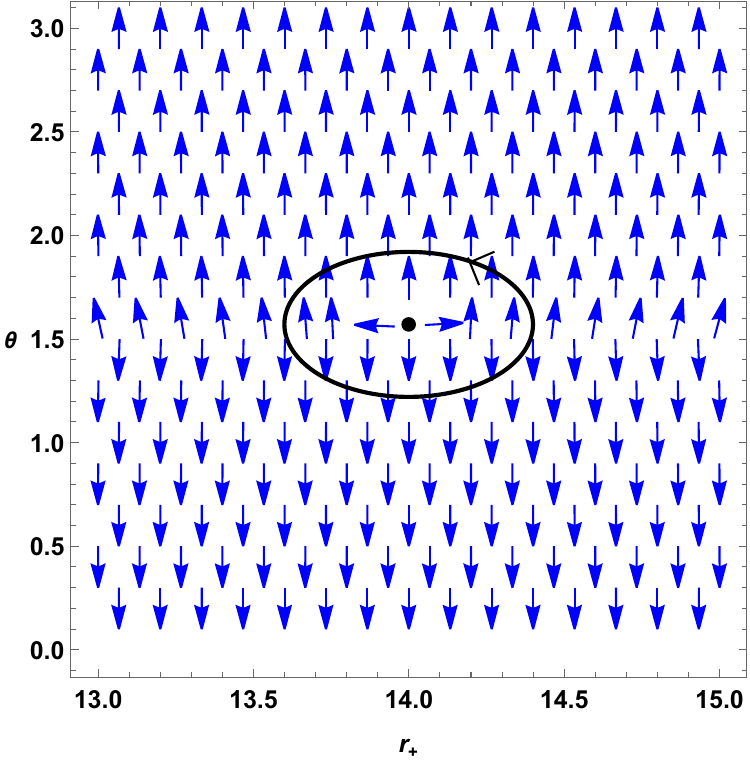}
\caption{}
\label{c3d}
\end{subfigure}
\begin{subfigure}{0.35\textwidth}
\includegraphics[width=\linewidth]{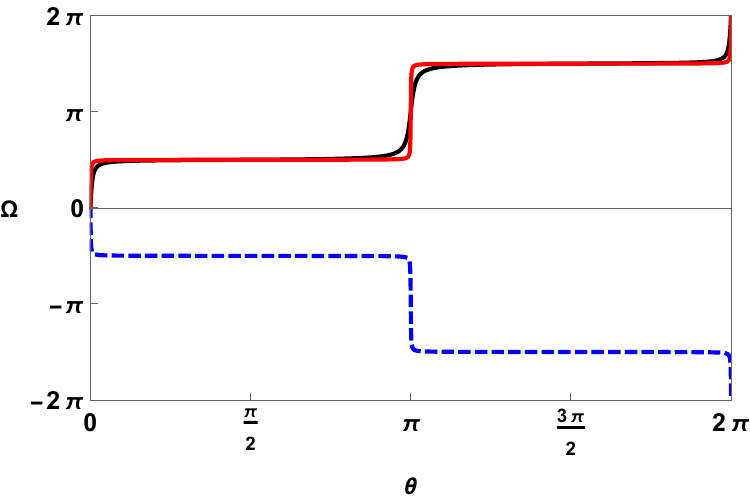}
\caption{}
\label{c3e}
\end{subfigure}
\caption{Fig. \protect\ref{c3b}, Fig. \protect\ref{c3c}, Fig. \protect\ref%
{c3d} shows vector plot for zero points $%
r_{+}=1.1478,r_{+}=3.7765,r_{+}=13.9994$ respectively. The winding number
calculations for zero points $r_{+}=1.1478,r_{+}=3.7765,r_{+}=13.9994$ are
represented by black solid line, blue dashed line, red solid line
respectively in Fig. \protect\ref{c3e}.}
\label{c3}
\end{figure}

If we consider the negative value of $R_{0}$, we observe a first-order phase
transition, as shown in Fig. \ref{c33}. Here, we plot $\tau$ vs $r_{+}$,
using the following values: $f_{R_{0}}=0.001$, $f(\varepsilon)=1.2$, $%
g(\varepsilon)=1.2$, $q=1$, $R_{0}=-0.1$, $\gamma=0.5$, and $k=1$. Fig. \ref%
{c3} clearly displays three black hole branches. The small black hole branch
is represented by the black solid line $(0\leq r_+ \leq 1.6579)$, the
intermediate black hole branch is represented by the blue dashed line $%
(1.6579 < r_+ \leq 7.4061)$, and the large black hole branch is represented
by the red solid line $(r_+ >1.6579)$. The winding numbers of these branches
are shown in Fig. \ref{c3e}. The winding number for the large and small
black hole branches is $+1$, while the intermediate black hole branch has a
winding number of $-1$. A positive winding number indicates a stable branch,
while a negative winding number indicates an unstable branch. The total
topological charge is $1-1+1=1$. 

It is observed that, apart from the sign of $R_{0}$, the topological charge
is independent of other thermodynamic parameters. In conclusion, for
topological black holes ($k=1$) in $F(R)$-ModMax gravity's rainbow in the
fixed charge ensemble, the topological charge is $0$ for positive values of $%
R_{0}$ and $+1$ for negative values of $R_{0}$.

\subsubsection{\textbf{Fixed potential (}$\protect\phi $\textbf{) ensemble}}

In this ensemble, the potential $\phi $ is kept fixed, which is conjugate to
charge $q$. The expression for $\phi $ is obtained in the following form 
\begin{equation}
\phi =\frac{\partial M}{\partial q}=\frac{qe^{-\gamma }f(\varepsilon )}{4\pi
g(\varepsilon )r_{+}},  \label{phi}
\end{equation}%
which is extracted from the equation (\ref{MM}).

The new mass in this ensemble is given by 
\begin{equation}
\tilde{M}=M-q\phi =\frac{(1+f_{R_{0}})\left( 12kg^{2}(\varepsilon
)-r_{+}^{2}R_{0}-\frac{192\pi ^{2}e^{\gamma }\phi ^{2}g^{4}(\varepsilon )}{%
(1+f_{R_{0}})}\right) r_{+}}{96\pi g^{3}(\varepsilon )f(\varepsilon )}.
\label{MMG}
\end{equation}

The free energy is obtained in the following form by using equations (\ref{S}%
) and (\ref{MMG}) follows 
\begin{eqnarray}
\mathcal{F} &=&\tilde{M}-\frac{S}{\tau }  \notag \\
&=&\frac{(1+f_{R_{0}})\left( kg^{2}(\varepsilon )\tau -2\pi g(\varepsilon
)f(\varepsilon )r_{+}-\frac{R_{0}\tau r_{+}^{2}}{12}\right) r_{+}}{8\pi
g^{3}(\varepsilon )f(\varepsilon )\tau }-\frac{2\pi e^{\gamma }\phi
^{2}g(\varepsilon )r_{+}}{f(\varepsilon )},
\end{eqnarray}%
by utilizing the equation $\phi ^{r}=\frac{\partial \mathcal{F}}{\partial
r_{+}}$ and the aforementioned relation, we can extract the component $\phi
^{r}$ from the vector field in the following form 
\begin{equation}
\phi ^{r}=\frac{g^{2}(\varepsilon )\left( k-\frac{16\pi ^{2}e^{\gamma }\phi
^{2}g^{2}(\varepsilon )}{(1+f_{R_{0}})}\right) \tau -4\pi g(\varepsilon
)f(\varepsilon )r_{+}-\frac{R_{0}\tau r_{+}^{2}}{4}}{\frac{8\pi
g^{3}(\varepsilon )f(\varepsilon )\tau }{(1+f_{R_{0}})}},
\end{equation}%
by considering the above equation, the zero points of the $\phi ^{r}$
component are obtained as 
\begin{equation}
\tau =\frac{4\pi g(\varepsilon )f(\varepsilon )r_{+}}{kg^{2}(\varepsilon )-%
\frac{r_{+}^{2}R_{0}}{4}-\frac{16\pi ^{2}e^{\gamma }\phi
^{2}g^{4}(\varepsilon )}{(1+f_{R_{0}})}}.
\end{equation}

\begin{figure}[h]
\centering
\begin{subfigure}{0.3\textwidth}
\includegraphics[height=5cm,width=5cm]{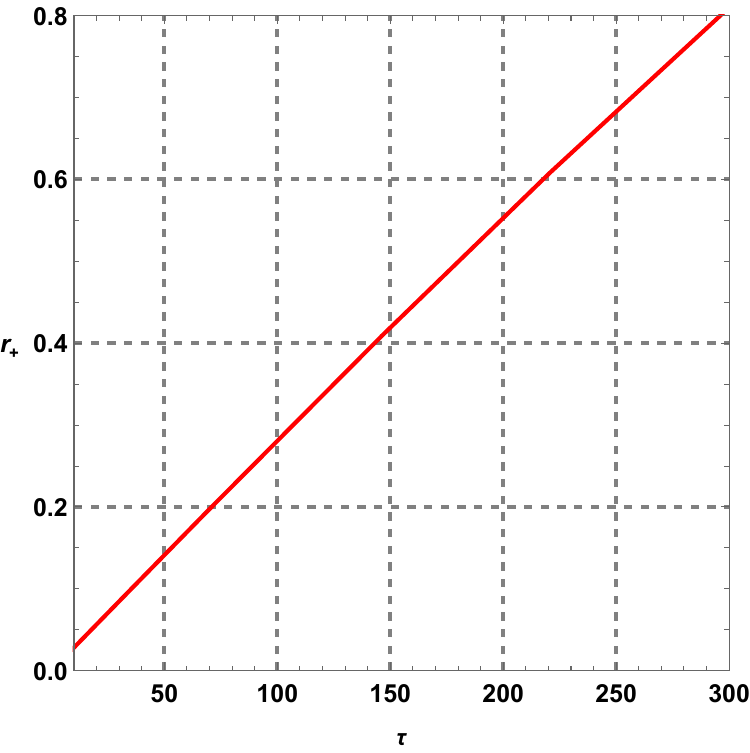}
\caption{}
\label{t3a}
\end{subfigure}
\begin{subfigure}{0.35\textwidth}
\includegraphics[height=5cm,width=5cm]{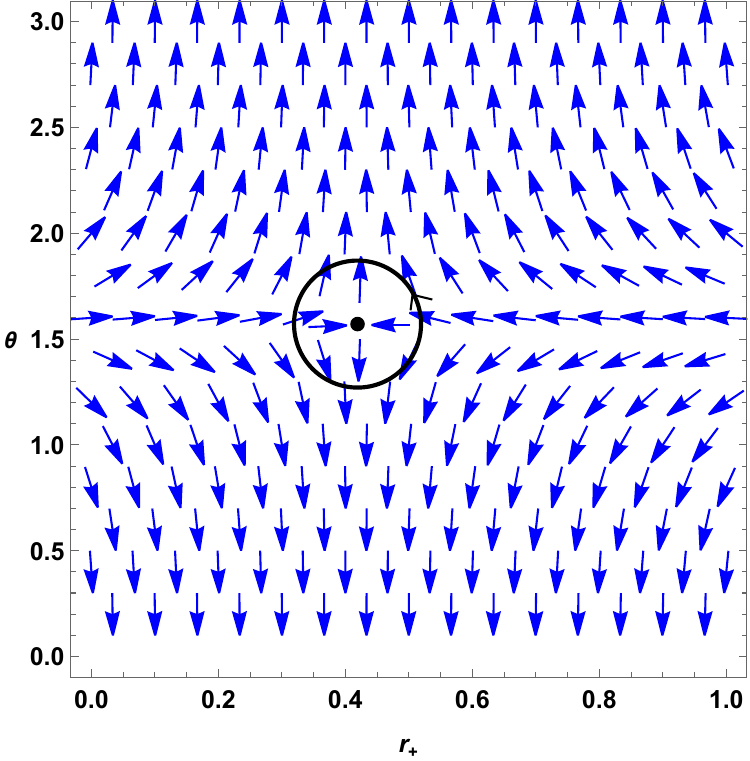}
\caption{}
\label{t3b}
\end{subfigure} 
\begin{subfigure}{0.3\textwidth}
\includegraphics[height=5cm,width=5cm]{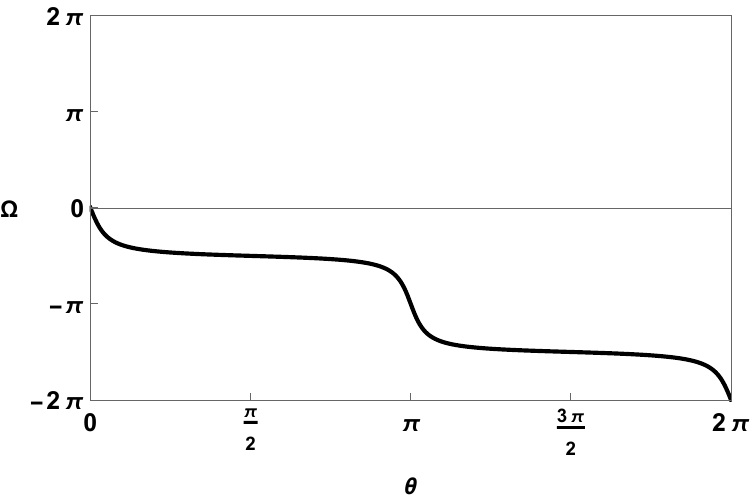}
\caption{}
\label{t3c}
\end{subfigure}
\caption{Topological charge of topological black holes $(k=1)$ in fixed
potential ensemble when positive values of $R_{0}$ is considered.}
\label{gc1}
\end{figure}

The plot in Fig. \ref{t3a} shows the relationship between $\tau$ and $r_{+}$%
. The parameter values used in this plot are as follows: $f_{R_{0}}=0.01$, $%
f(\varepsilon)=1.1$, $g(\varepsilon)=1.1$, $\phi=0.05$, $R_{0}=0.001$, $%
\gamma=0.5$, and $k=1$. Only one branch of the black hole is observed. In
Fig. \ref{t3b}, a vector plot is displayed for the components $\phi^{r}$ and 
$\phi^{\theta}$ with $\tau=150$. The zero points of the vector field are
located at $r_{+}=0.41916$. Fig. \ref{t3c} shows that the topological charge
corresponding to $r_{+}=0.41916$ is $-1$, as indicated by the black line.

\begin{figure}[h!]
\centering
\begin{subfigure}{0.3\textwidth}
\includegraphics[width=\linewidth]{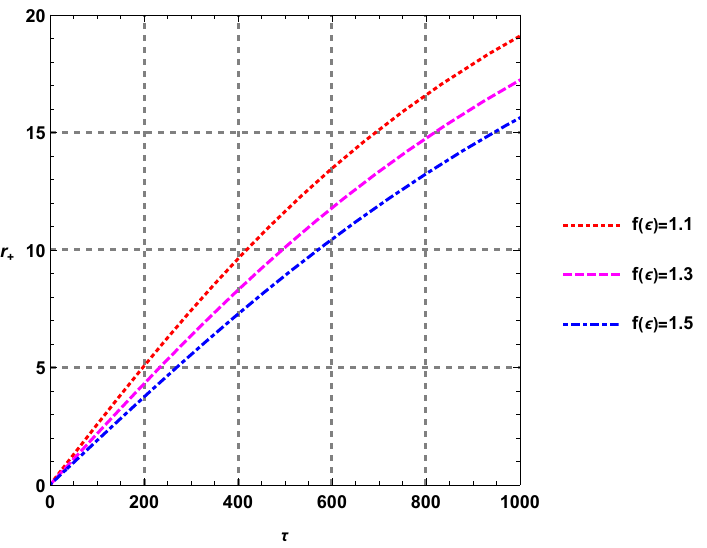}
\caption{}
\label{t4a}
\end{subfigure}
\begin{subfigure}{0.3\textwidth}
\includegraphics[width=\linewidth]{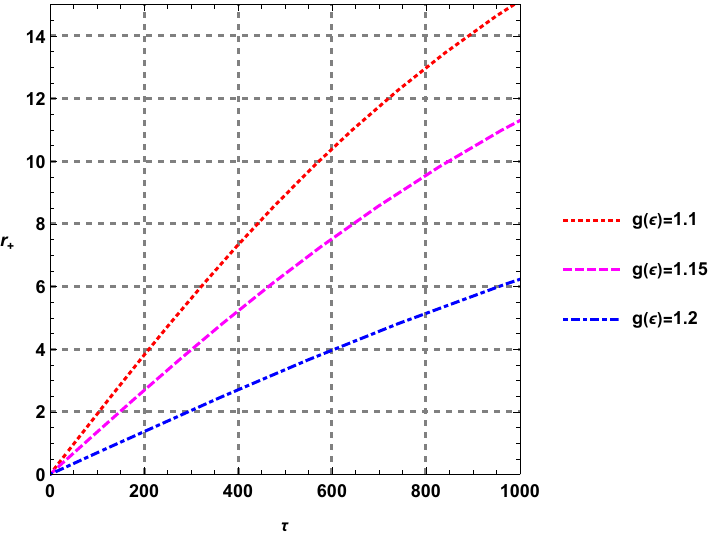}
\caption{}
\label{t4b}
\end{subfigure} \hspace{0.6cm} 
\begin{subfigure}{0.3\textwidth}
\includegraphics[width=\linewidth]{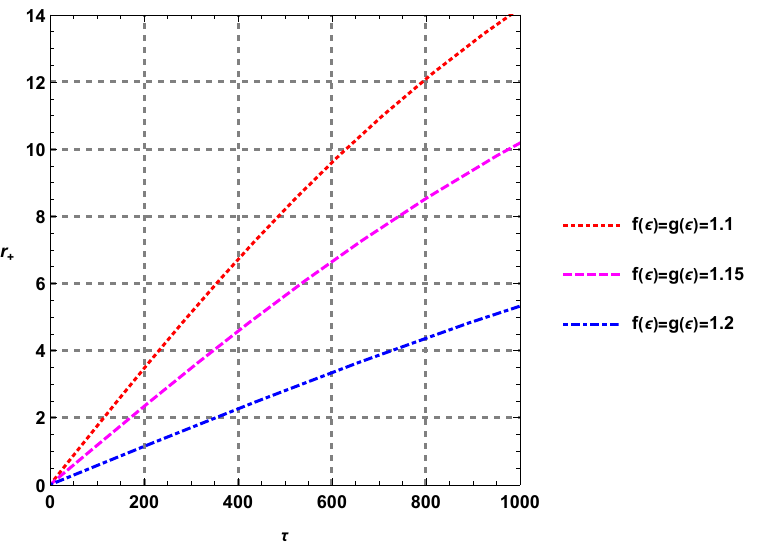}
\caption{}
\label{t4c}
\end{subfigure}
\begin{subfigure}{0.3\textwidth}
\includegraphics[width=\linewidth]{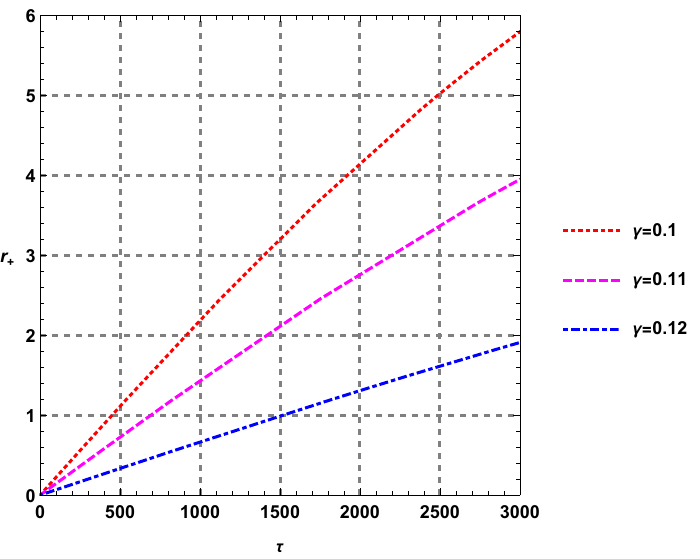}
\caption{}
\label{t4e}
\end{subfigure}
\begin{subfigure}{0.3\textwidth}
\includegraphics[width=\linewidth]{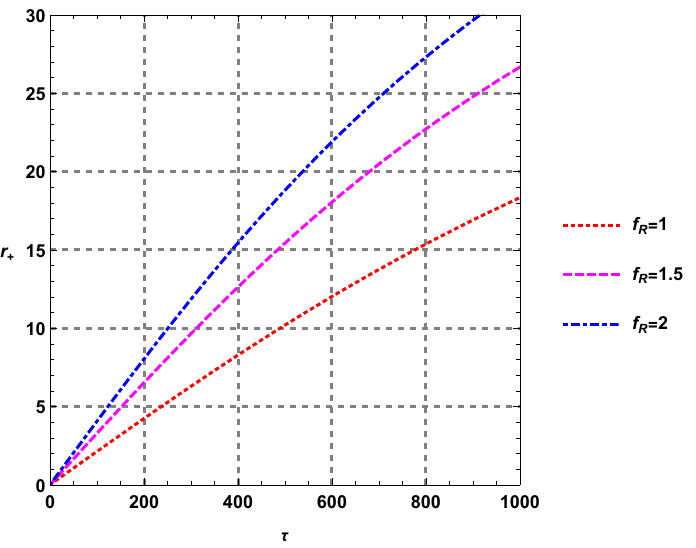}
\caption{}
\label{t4d}
\end{subfigure}
\caption{$\protect\tau$ vs $r_+$ of topological black holes $(k=1)$ in fixed
potential ensemble when positive value of $R_0$ is considered.}
\label{gc2s}
\end{figure}

Similar to the fixed $q$ ensemble, the topological charge in this ensemble
also remains unchanged when the rainbow function, represented by $%
g(\varepsilon)$, $f(\varepsilon)$, $\gamma$, and $f_{R_{0}}$, varies.
However, the values of $\phi$, $R_{0}$, and $k$ are fixed at $0.05$, $0.001$%
, and $1$ respectively. Figure. \ref{t4a} illustrates the variation of $%
f(\varepsilon)$ when $g(\varepsilon)$ is held constant at $1$, along with $%
\gamma$ at $0.5$ and $f_{R_{0}}$ at $0.01$. Similarly, in Figure. \ref{t4b},
the change in $\tau$ versus $r_{+}$ is shown when $g(\varepsilon)$ is varied
while keeping $f(\varepsilon)$ at $1$, $\gamma$ at $0.5$, and $f_{R_{0}}$ at 
$0.01$. Additionally, Figure. \ref{t4c} demonstrates the impact of both $%
f(\varepsilon)$ and $g(\varepsilon)$ on the $\tau$ versus $r_{+}$ curve,
while maintaining $\gamma$ at $0.5$ and $f_{R_{0}}$ at $0.01$. The variation
of $\gamma$ is depicted in Figure. \ref{t4c}, with fixed values of $%
f(\varepsilon)$ at $1.5$, $g(\varepsilon)$ at $1.5$, and $f_{R_{0}}$ at $%
0.01 $. Finally, Figure \ref{t4d} displays the variation of $f_{R_{0}}$,
while holding constant values for $f(\varepsilon)$. Figure. \ref{gc2s}
indicates that due to the change in the ensemble, the topological charge
shifts to $-1$, although it remains independent of thermodynamic quantities
in this ensemble as well. 

\begin{figure}[h]
\centering
\begin{subfigure}{0.35\textwidth}
\includegraphics[width=\linewidth]{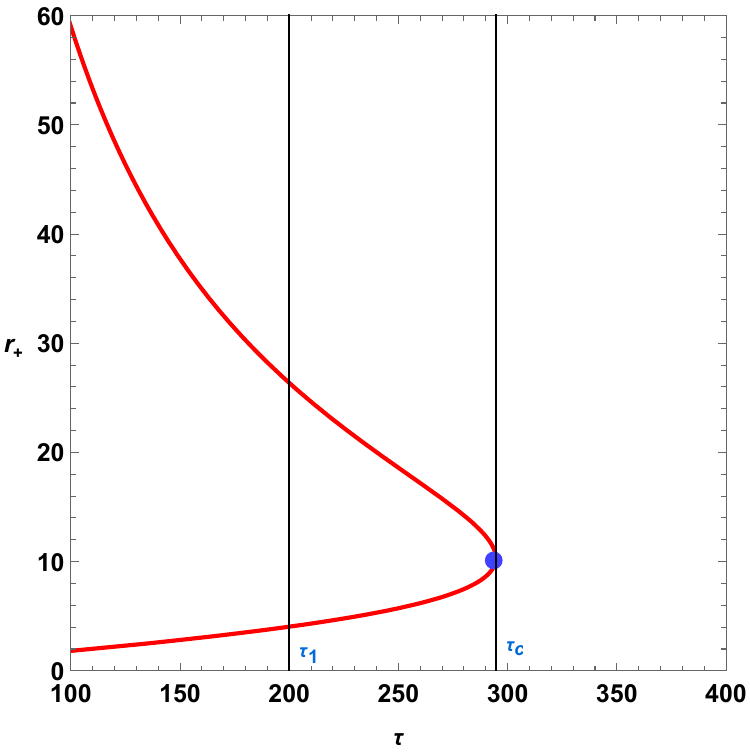}
\caption{}
\label{gc3a}
\end{subfigure}
\begin{subfigure}{0.35\textwidth}
\includegraphics[width=\linewidth]{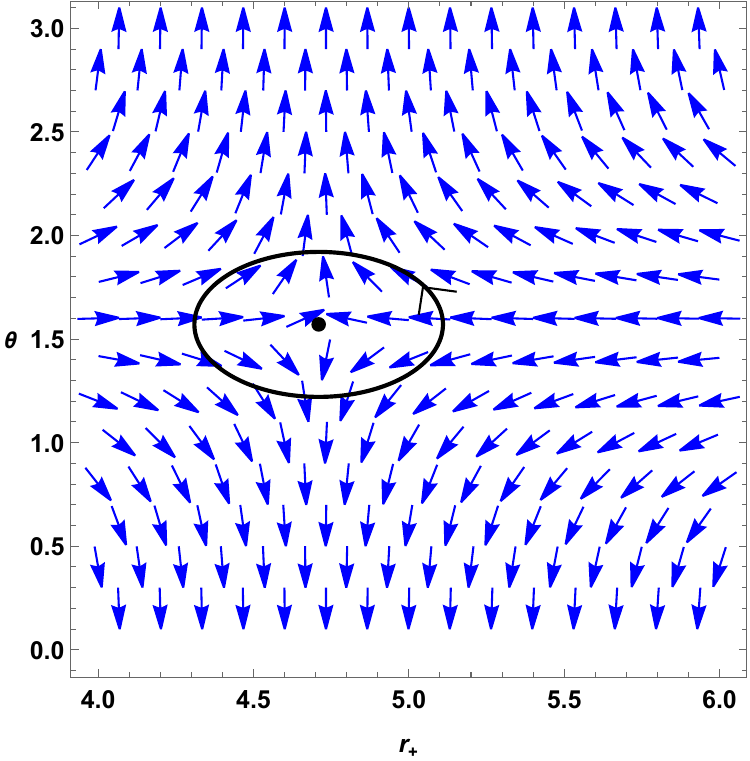}
\caption{}
\label{gc3b}
\end{subfigure} 
\begin{subfigure}{0.35\textwidth}
\includegraphics[width=\linewidth]{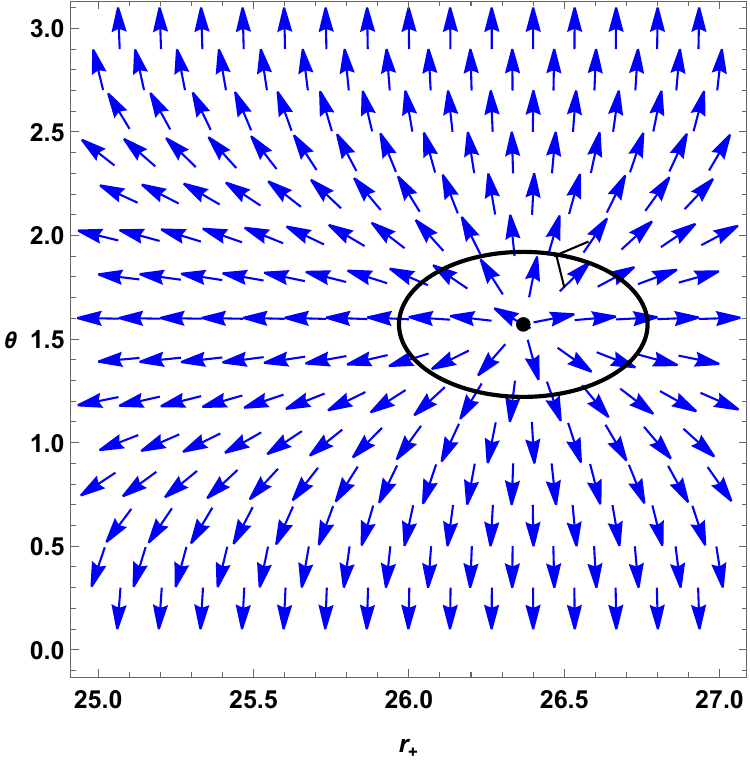}
\caption{}
\label{gc3c}
\end{subfigure} 
\begin{subfigure}{0.35\textwidth}
\includegraphics[width=\linewidth]{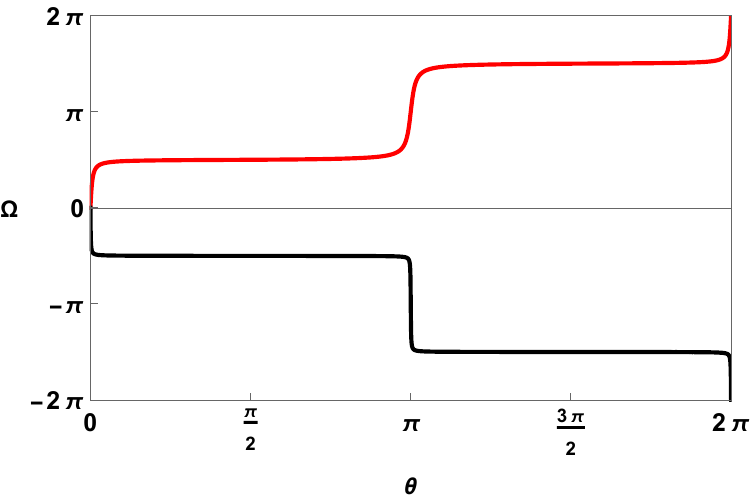}
\caption{}
\label{gc3d}
\end{subfigure}
\caption{Topological charge of topological black holes $(k=1)$ in fixed
potential ensemble when negative values of $R_0$ is considered. In Fig. 
\protect\ref{gc3b} and Fig. \protect\ref{gc3c}, vector plots are shown for $%
f(\protect\varepsilon)=g(\protect\varepsilon)=1.7$ case. In Fig. \protect\ref%
{gc3c} the red solid line and black solid represents winding number
calculation around the zero points $r_+=4.042$ and $r_+=26.382$,
respectively. }
\label{gc3}
\end{figure}

When negative values of $R_{0}$ are considered, the topological charge
changes to $0$, as shown in Fig. \ref{gc3}. We have calculated the
topological charge for the red-colored solid line in Fig. \ref{gc3a}. The
values used for calculations are: $f_{R_{0}}=0.01$, $f(\varepsilon)=1.1$, $%
g(\varepsilon)=1.1$, $\phi=0.05$, $R_{0}=-0.01$, $\gamma =0.5$, and $k=1$. A
generation is observed at $r_{+}=10.324$, represented by the blue dot in
Fig. \ref{gc3a}. For $\tau_{1}=200$, the zero points are calculated at $%
r_{+}=4.042$ and $r_{+}=26.382$, as shown in Fig. \ref{gc3b} and Fig. \ref%
{gc3c}, respectively. In Fig. \ref{gc3d}, the topological charge
calculations for the zero points $r_{+}=4.042$ and $r_{+}=26.382$ are shown
as black-colored and red-colored solid lines. It can be observed that the
topological charge is $0$. 

\begin{figure}[h!]
\centering
\begin{subfigure}{0.35\textwidth}
\includegraphics[width=\linewidth]{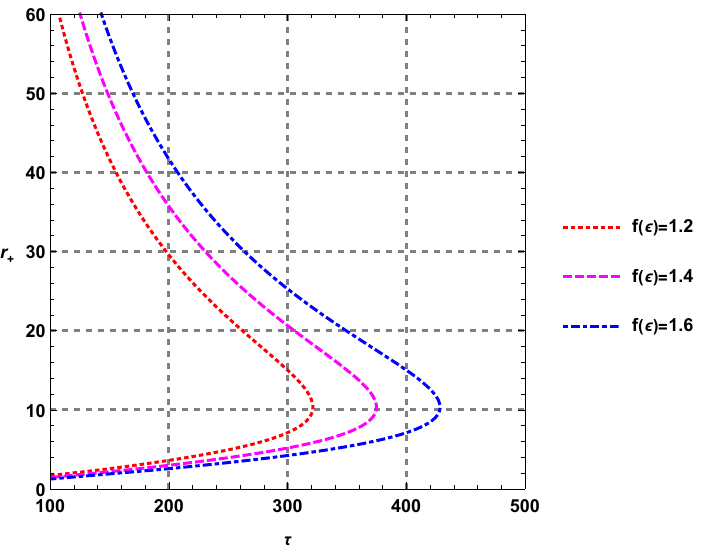}
\caption{}
\label{gc2a}
\end{subfigure}
\begin{subfigure}{0.35\textwidth}
\includegraphics[width=\linewidth]{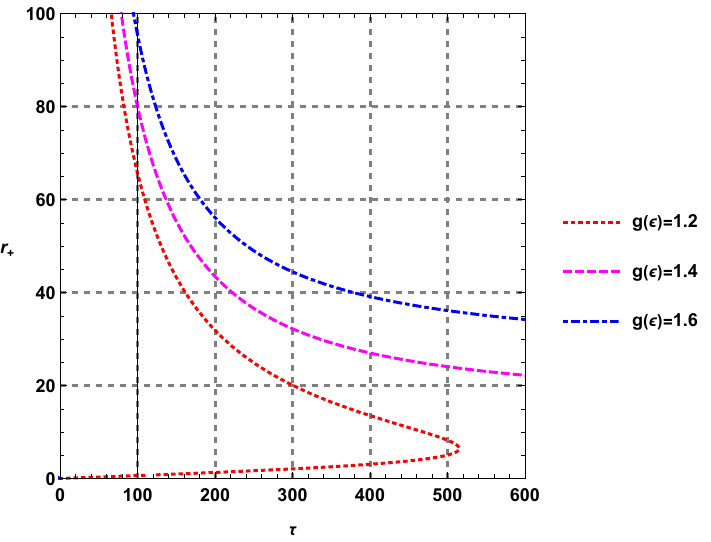}
\caption{}
\label{gc2b}
\end{subfigure} \hspace{0.6cm} 
\begin{subfigure}{0.35\textwidth}
\includegraphics[width=\linewidth]{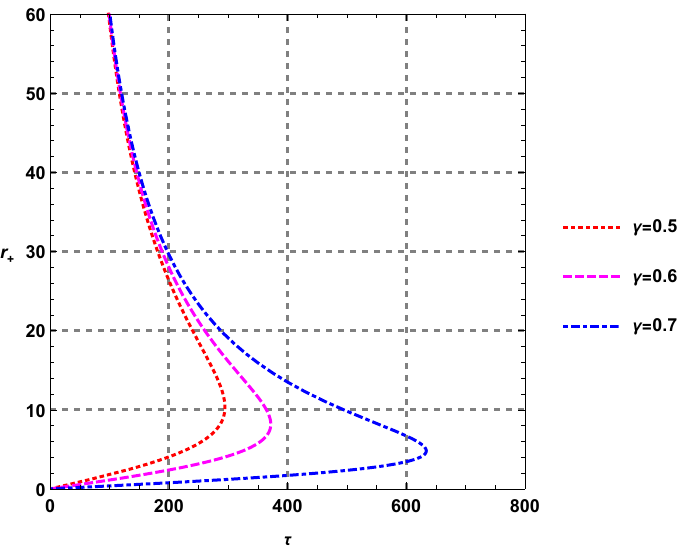}
\caption{}
\label{gc2c}
\end{subfigure}
\begin{subfigure}{0.35\textwidth}
\includegraphics[width=\linewidth]{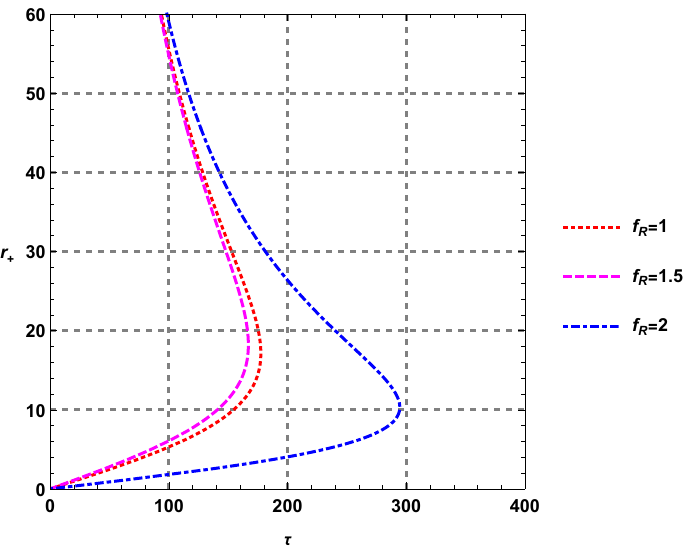}
\caption{}
\label{gc2d}
\end{subfigure}
\caption{$\protect\tau$ vs $r_+$ of topological black holes $(k=1)$ in fixed
potential ensemble when positive value of $R_0$ is considered.}
\label{gc2}
\end{figure}

In Fig. \ref{gc2}, the graphs show the dependence of the nature of $\tau$ vs 
$r_{+}$ on thermodynamic parameters. Interestingly, in Fig. \ref{gc2b}, a
change in topological charge from $0$ to $1$ is observed when the parameter $%
g(\varepsilon)$ is varied while keeping the other parameters fixed. For
instance, in Fig. \ref{gc2b}, the magenta dashed line represents the $\tau$
vs $r_{+}$ graph for $f_{R}=0.01, f(\varepsilon)=1.6, g(\varepsilon)=141,
\phi=0.05, R_{0}=-0.01$, and $\gamma=0.5$. Here, a single black hole branch
with a topological charge of $+1$ is obtained. Therefore, when $R_{0}$ is
negative, the grand canonical ensemble yields two types of topological
charge, namely $0$ and $+1$, depending on the value of $g(\varepsilon)$. 

In conclusion, the topological charge for topological black holes with $k=1$
in $F(R)$-ModMax gravity's rainbow in the fixed potential ensemble is $-1$
when a positive value of $R_{0}$ is taken. Similarly, when negative values
of $R_{0}$ are considered, the topological charge is found to be $0$ or $+1$%
. The results in both ensembles are summarized in Table \ref{table1}. 
\begin{table}[h]
\caption{Our results in two ensembles.}
\label{table1}\centering
\par
\begin{tabular}{ccc}
\hline\hline
& Fixed charge $(q)$ ensemble \hspace{0.3cm} & Fixed potential $(\phi)$
ensemble \\ \hline\hline
Topological Charge & 0 or 1 & -1, 0 or +1 \\ 
Generation Point & 0 or 1 & 1 or 0 \\ 
Annihilation Point & 1 & 0 \\ \hline\hline
\end{tabular}%
\end{table}

\subsection{\textbf{For flat (}$k=0$\textbf{) curvature hypersurface}}

For topological black holes with boundary of $t=$ constant and $r=$constant
having flat curvature hypersurface, equation for $\tau $ becomes 
\begin{eqnarray}
\tau _{fixedq} &=&-\frac{4\pi g(\varepsilon )f(\varepsilon )r_{+}^{3}}{\frac{%
q^{2}e^{-\gamma }g^{2}(\varepsilon )f^{2}(\varepsilon )}{(1+f_{R_{0}})}+%
\frac{r_{+}^{4}R_{0}}{4}}, \\
&&  \notag \\
\tau _{fixed\phi } &=&-\frac{4\pi g(\varepsilon )f(\varepsilon )r_{+}}{\frac{%
16\pi ^{2}e^{\gamma }g^{4}(\varepsilon )\phi ^{2}}{(1+f_{R_{0}})}+\frac{%
r_{+}^{2}R_{0}}{4}}.
\end{eqnarray}

\begin{figure}[h]
\centering
\begin{subfigure}{0.3\textwidth}
\includegraphics[width=\linewidth]{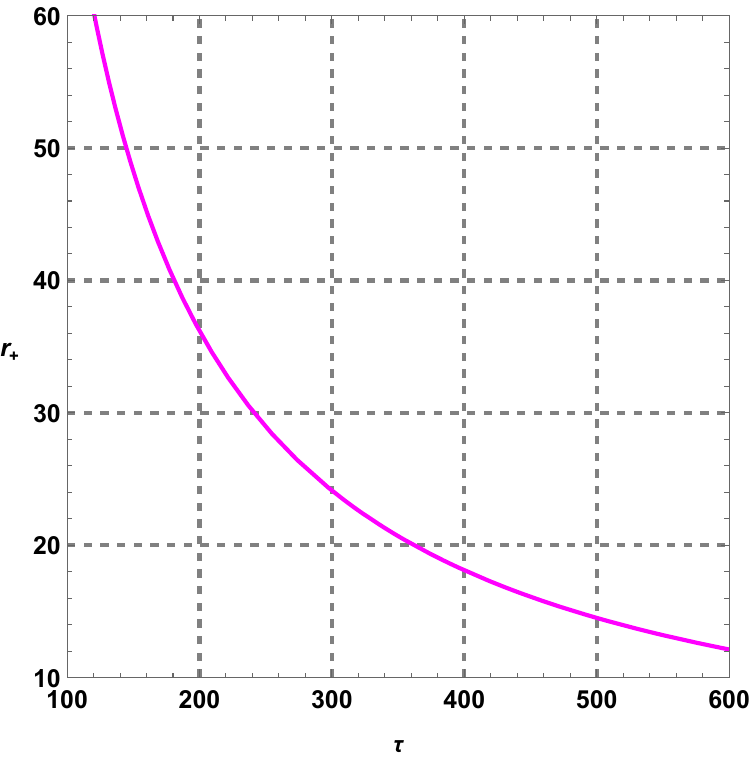}
\caption{}
\label{t5a}
\end{subfigure}
\begin{subfigure}{0.3\textwidth}
\includegraphics[width=\linewidth]{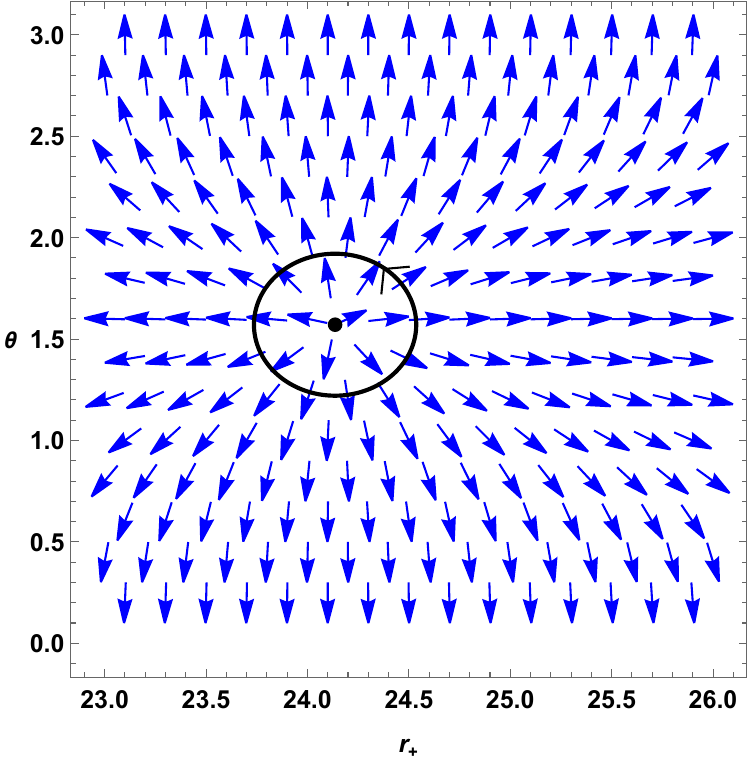}
\caption{}
\label{t5b}
\end{subfigure}
\begin{subfigure}{0.3\textwidth}
\includegraphics[width=\linewidth]{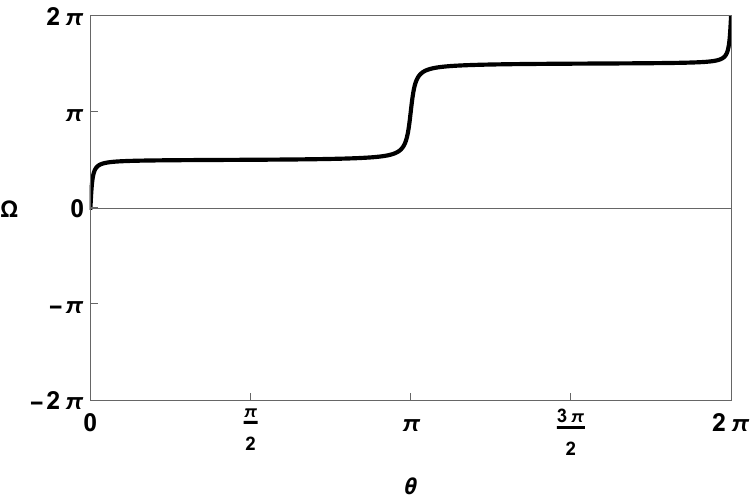}
\caption{}
\label{t5c}
\end{subfigure}
\caption{ Plots for topological black holes $(k=0)$ in fixed charge ensemble
at $f_{R_{0}}=0.01, f(\protect\varepsilon )=1.2, g(\protect\varepsilon)=1.2,
q=0.5, R_{0}=-0.01, \protect\gamma =0.5$ and $k=0$. Fig. $\left( a\right)$
shows $\protect\tau $ vs $r_{+}$ plot, Fig. $\left( b\right)$ is the plot of
vector field $n$ on aportion of $r_{+}-\protect\theta $ plane for $\protect%
\tau =300$. The zero points is located at $r_{+}=24.1362$. In Fig. $\left(
c\right)$, computation of the contours around the zero points $r_{+}=24.1362$
is shown in black colored solid line.}
\label{t5}
\end{figure}

\begin{figure}[h]
\centering
\begin{subfigure}{0.3\textwidth}
\includegraphics[width=\linewidth]{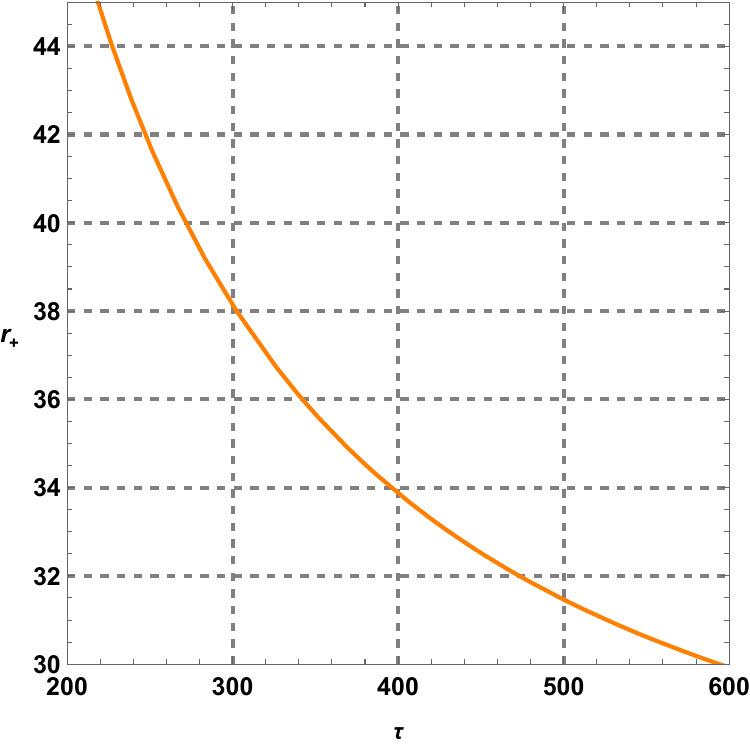}
\caption{}
\label{t6a}
\end{subfigure}
\begin{subfigure}{0.3\textwidth}
\includegraphics[width=\linewidth]{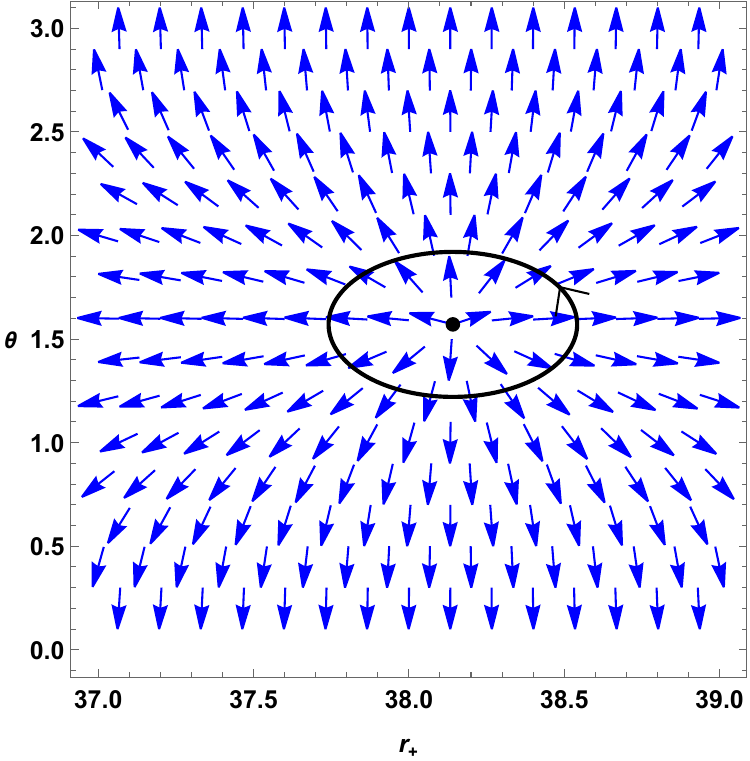}
\caption{}
\label{t6b}
\end{subfigure}
\begin{subfigure}{0.3\textwidth}
\includegraphics[width=\linewidth]{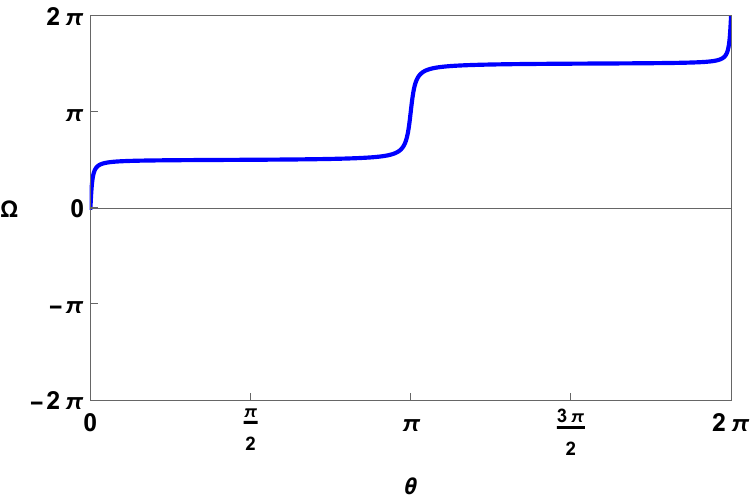}
\caption{}
\label{t6c}
\end{subfigure}
\caption{ Plots for topological black holes $(k=0)$ in fixed potential
ensemble at $f_{R_{0}}=0.01, f(\protect\varepsilon )=1.2, g(\protect%
\varepsilon )=1.2, \protect\phi =0.05, R_{0}=-0.01, \protect\gamma =0.5$ and 
$k=0$. Figure $\left( a\right) $ shows $\protect\tau $ vs $r_{+}$ plot,
figure $\left( b\right) $ is the plot of vector field $n$ on a portion of $%
r_{+}-\protect\theta $ plane for $\protect\tau=300$. The zero points is
located at $r_{+}=38.1417$. In figure $\left(c\right) $, computation of the
contours around the zero points $r_{+}=38.1417$ is shown in blue colored
solid line.}
\label{t6}
\end{figure}

It is clear from both the equations that $\tau $ is always negative when $%
R_{0}$ is positive. Therefore, to have positive value of $\tau $, the scalar
curvature $R_{0}$, must be negative in both ensembles. Fig. \ref{t5} and
Fig. \ref{t6} show that the topological charge for the topological black
hole with a flat curvature hypersurface in $F(R)$-ModMax gravity's rainbow
is $1$ for both ensembles.


\subsection{\textbf{For hyperbolic (}$k=-1$\textbf{) curvature hypersurface}}

For topological black holes with a boundary of constant $t$ and constant $r$%
, characterized by a hypersurface with hyperbolic curvature, the equation
for $\tau $ can be expressed as 
\begin{eqnarray}
\tau _{fixedq} &=&\frac{-4\pi g(\varepsilon )f(\varepsilon )r_{+}}{%
g^{2}(\varepsilon )+\frac{q^{2}e^{-\gamma }g^{2}(\varepsilon
)f^{2}(\varepsilon )}{(1+f_{R_{0}})r_{+}^{2}}+\frac{R_{0}r_{+}^{2}}{4}}, \\
&&  \notag \\
\tau _{fixed\phi } &=&\frac{-4\pi g(\varepsilon )f(\varepsilon )r_{+}}{%
g^{2}(\varepsilon )+\frac{16\pi ^{2}e^{-\gamma }\phi ^{2}g^{4}(\varepsilon )%
}{(1+f_{R_{0}})}+\frac{R_{0}r_{+}^{2}}{4}}.
\end{eqnarray}

\begin{figure}[h]
\centering
\begin{subfigure}{0.3\textwidth}
\includegraphics[width=\linewidth]{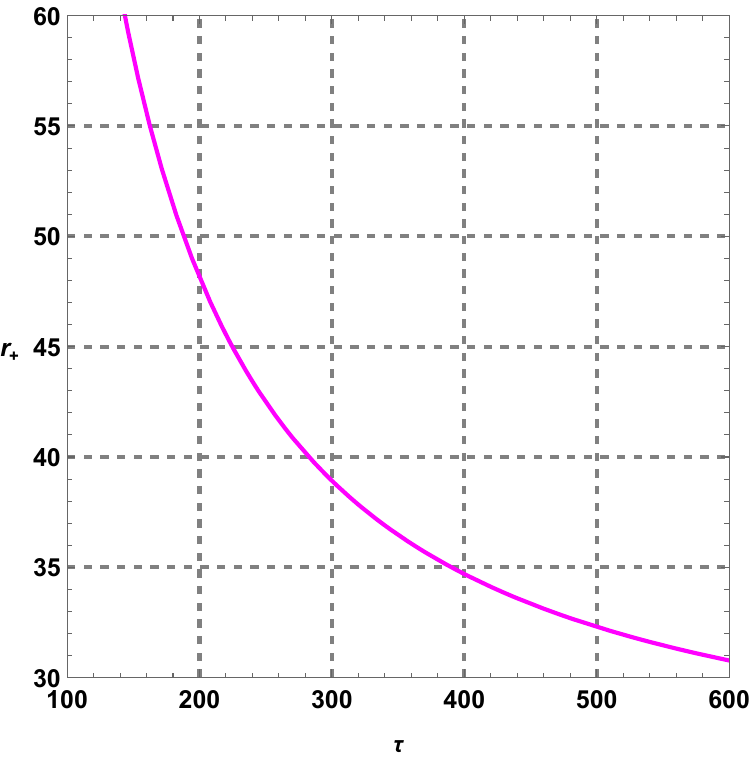}
\caption{}
\label{t7a}
\end{subfigure}
\begin{subfigure}{0.3\textwidth}
\includegraphics[width=\linewidth]{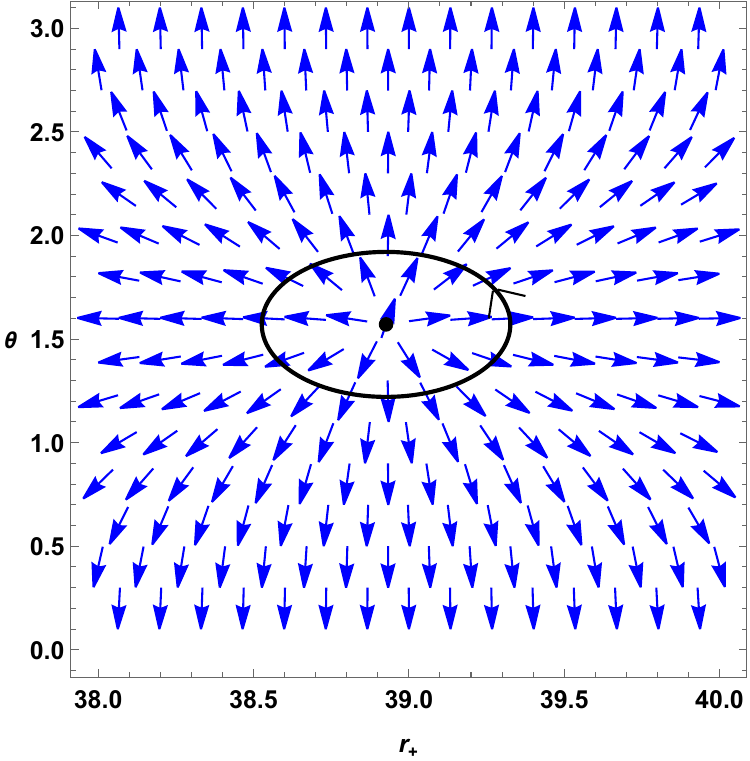}
\caption{}
\label{t7b}
\end{subfigure}
\begin{subfigure}{0.3\textwidth}
\includegraphics[width=\linewidth]{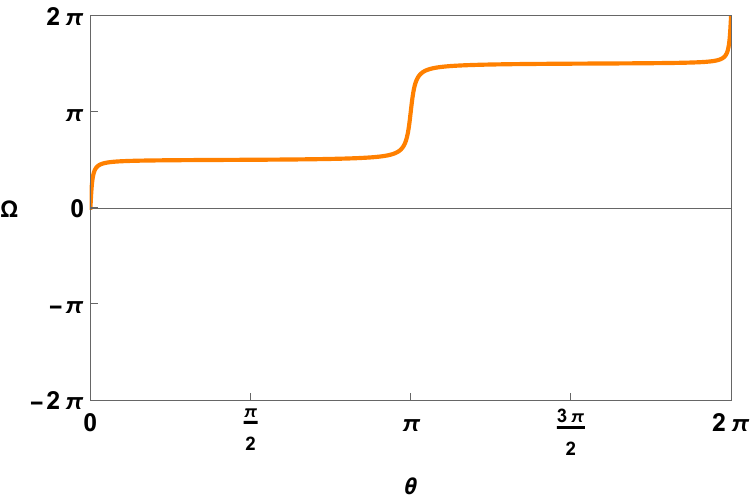}
\caption{}
\label{t7c}
\end{subfigure}
\caption{Plots for topological black holes $(k=-1)$ in fixed charge ensemble
at $f_{R_{0}}=0.01, f(\protect\varepsilon )=1.2, g(\protect\varepsilon)=1.2,
q=0.5, R_{0}=-0.01, \protect\gamma =0.5$ and $k=-1$. Fig. $\left( a\right)$
shows $\protect\tau $ vs $r_{+}$ plot, Fig. $\left( b\right) $ is the plot
of vector field $n$ on a portion of $r_{+}-\protect\theta $ plane for $%
\protect\tau =300$ . The zero points is located at $r_{+}=38.9266$. In Fig. $%
\left( c\right) $, computation of the contours around the zero points $%
r_{+}=38.9266$ is shown in orange colored solid line.}
\label{t7}
\end{figure}

\begin{figure}[h]
\centering
\begin{subfigure}{0.3\textwidth}
\includegraphics[width=\linewidth]{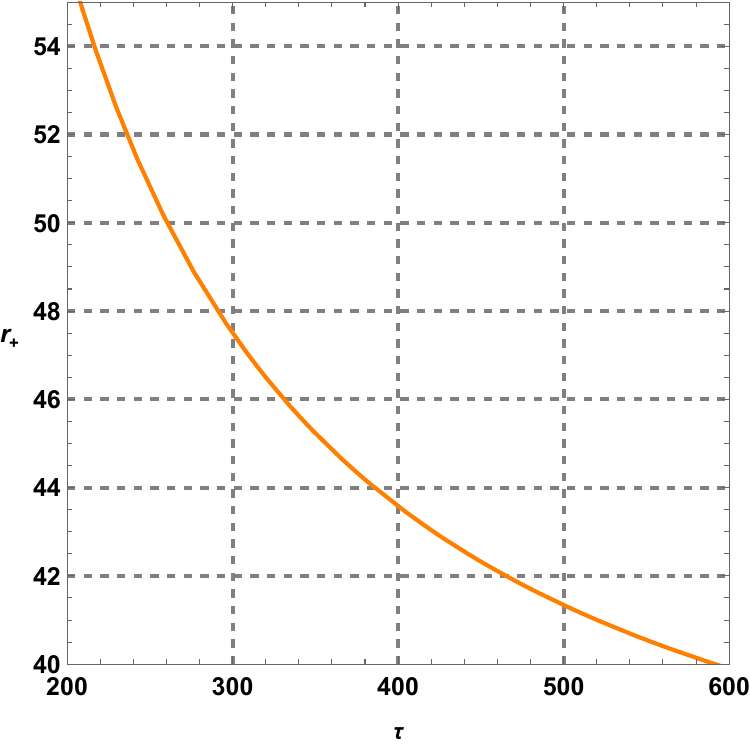}
\caption{}
\label{t8a}
\end{subfigure}
\begin{subfigure}{0.3\textwidth}
\includegraphics[width=\linewidth]{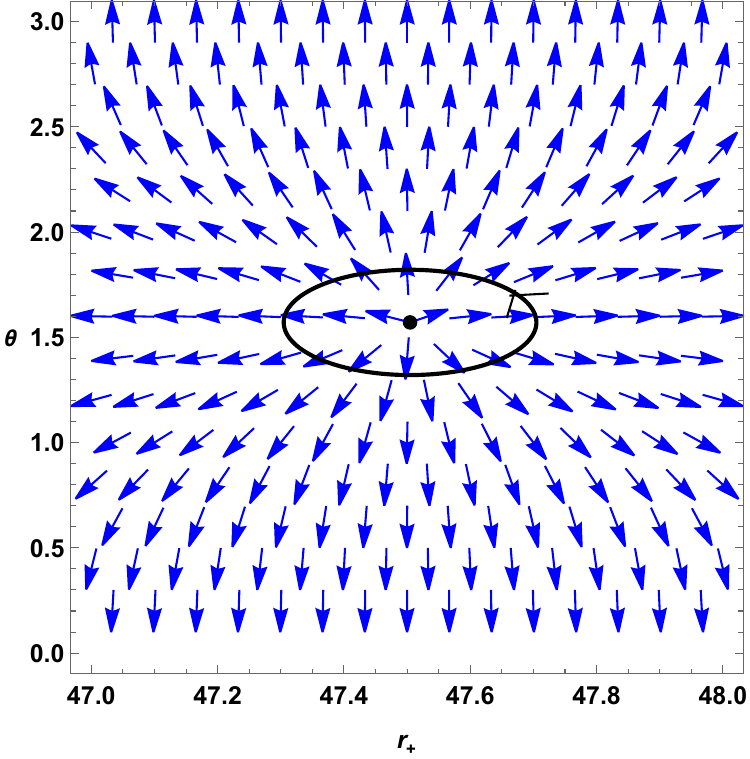}
\caption{}
\label{t8b}
\end{subfigure}
\begin{subfigure}{0.3\textwidth}
\includegraphics[width=\linewidth]{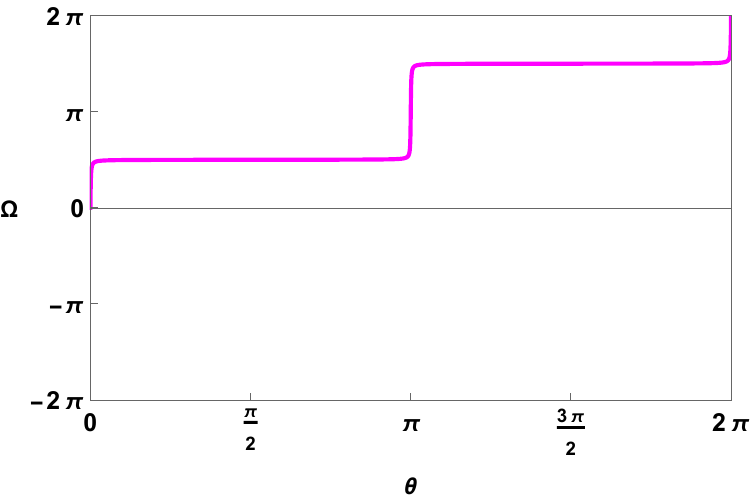}
\caption{}
\label{t8c}
\end{subfigure}
\caption{ Plots for topological black holes $(k=-1)$ in fixed potential
ensemble at $f_{R_{0}}=0.01, f(\protect\varepsilon )=1.2, g(\protect%
\varepsilon )=1.2, \protect\phi=0.05, R_{0}=-0.01, \protect\gamma =0.5$ and $%
k=-1$. Fig. $\left(a\right) $ shows $\protect\tau $ vs $r_{+}$ plot, Fig. $%
\left( b\right)$ is the plot of vector field $n$ on a portion of $r_{+}-%
\protect\theta $ plane for $\protect\tau=300$. The zero points is located at 
$r_{+}=47.5046$. In Fig. $\left( c\right)$, computation of the contours
around the zero points $r_{+}=47.5046$ is shown in magenta colored solid
line. }
\label{t8}
\end{figure}

Due to the positive temperature condition, $R_{0}$ must be negative.
Figures. \ref{t7} and \ref{t8} show the topological charge for the
topological black hole with hyperbolic curvature hypersurface in $F(R)$%
-ModMax gravity's rainbow is $1$ for both ensembles.


\section{\textbf{Conclusion}}

In this paper, we investigated the effects of high energy and topological
parameters on black holes in $F(R)$ gravity, simultaneously. To achieve this
objective, we took into account two corrections to $F(R)$ gravity: i)
energy-dependent spacetime with different topological constants, and ii) a
nonlinear electrodynamics field. In other words, we combined $F(R)$
gravity's rainbow with different topological constants and the ModMax
nonlinear electrodynamics theory to examine the impact of high energy and
topological parameters on the physics of black holes. To begin with, we
first extracted exact solutions in $F(R)$-ModMax gravity's rainbow. To
determine the singularities of the solutions obtained in the
energy-dependent spacetime with different topological constants, we examined
the Kretschmann scalar. Our calculations revealed the presence of an
essential curvature singularity at the coordinate $r=0$. To analyze the
asymptotical behavior of energy-dependent spacetime with different
topological constants in the $F(R)$-ModMax theory of gravity, we examined
the Kretschmann scalar and the metric function. Our findings indicated that
the energy-dependent spacetime is the asymptotically (A)dS when $%
R_{0}=4\Lambda $, and $\Lambda >0$ ($\Lambda <0$). Moreover, this
asymptotical behavior was independent of $\gamma $ and $k$. In other words,
the parameters of ModMax and the topological constant did not affect the
asymptotical behavior of the energy-dependent spacetime, but it depended on
the rainbow function $g\left( \varepsilon \right)$. The system we studied
had various parameters that affected the event horizon. It has been observed
that larger black holes are associated with a negative topological constant (%
$k=-1$). These black holes have high values for mass, electric charge, $%
\gamma$, $f_{R_{0}}$, and $g(\varepsilon)$ but a lower $f(\varepsilon)$.
Furthermore, our research has shown that changing specific parameter values
makes the radius of the event horizon more sensitive to $k=+1$ (except when $%
g\left(\varepsilon\right)$ increases, see Fig. \ref{Fig3}) when compared to
other values of the topological constant.

In section \ref{Thermodynamics}, we calculated the conserved and
thermodynamic quantities of the topological black hole solutions in $F(R)$
gravity's rainbow to check the first law of thermodynamics. We determined
the Hawking temperature, electrical charge, electrical potential, entropy,
and total mass of these black holes. Subsequently, we confirmed that these
thermodynamic quantities satisfy the first law of thermodynamics. By
calculating the Hawking temperature, we discovered that only the larger
black holes could have a positive temperature when $R_{0}<0$. Furthermore,
we found that the temperature had a single real root. The temperature was
negative before this root but became positive afterwards. Additionally, our
findings regarding the mass of these black holes revealed that for $k=+1$
and $k=0$, this quantity was always positive. However, for $k=-1$, there
were two roots at which the mass of large and small black holes was
positive. Indeed, we encountered the positive mass of these black holes in
the range $r_{+}<r_{1_{M=0}}$ and $r_{+}>r_{2_{M=0}}$ for $k=-1$. It was
negative in the range $r_{1_{M=0}}<r_{+}<r_{2_{M=0}}$. This imposed the
existence of two roots for the mass when $k=-1$.

In section \ref{tt}, thermodynamic topology is investigated in two
ensembles. We used the off-shell free energy method, where black holes are
assumed to be defects in their thermodynamic spaces. We studied the local
and global topology of these black holes by computing the topological
charges of the defects using Duan's $\phi$ mapping technique. It is observed
that the topological charge for topological black holes with the boundary of 
$t=$ constant and $r=$ constant having elliptic ($k=1$ curvature
hypersurface in $F(R)$-ModMax gravity's rainbow is $1$ or $0$ in the fixed
charge ensemble and $-1$ or $0$ in the fixed potential ensemble depending
upon the sign of $R_0$. On the other hand, black holes with the boundary of $%
t=$ constant and $r=$constant having flat ($k=0$) or hyperbolic ($k=-1$)
curvature hypersurface has a topological charge $1$ in both the ensemble. It
is seen that, in a fixed charge ensemble, the topological charge is
independent of variations in the respective thermodynamic parameters, except
for the sign of $R_0.$ Moreover, in fixed $\phi$ ensemble, the topological
charge is found to be $0$, $-1$ or $+1$ for black holes with the boundary of 
$t=$constant and $r=$constant having elliptic ($k=1$ curvature hypersurface.
We observe that the topological classes of these black holes are dependent
on the value of the rainbow function, the sign of the scalar curvature and
the choice of ensembles. On the other hand in fixed $\phi$ ensemble, the
topological charge is again found to be $+1$ for black holes with the
boundary of $t=$ constant and $r=$constant having flat ($k=0$) or hyperbolic
($k=-1$) curvature hypersurface. Hence, we can conclude that based on the
topological charge, topological black holes in $F(R)$-ModMax gravity's
rainbow can be classified into three topological classes: $-1$, $0$, and $+1$%
. This classification depends on the choice of ensemble, the value of the
topological parameter $K$, and the values of the thermodynamic parameters
specific to the chosen ensemble.


\section{Acknowledgments}

BEP would like to thank the University of Mazandaran. BH would like to thank
DST-INSPIRE, Ministry of Science and Technology fellowship program, Govt. of
India for awarding the DST/INSPIRE Fellowship[IF220255] for financial
support.

\end{document}